\documentclass[twocolumn,showpacs,preprintnumbers,prb,aps]{revtex4-1}
\usepackage{hyperref}
\pdfoutput=1

\usepackage[pdftex]{graphicx}
\usepackage{dcolumn}
\usepackage{amsfonts,amsmath,amssymb,bm}
\usepackage{etoolbox} 
\usepackage{subfigure}

\begin{document}

\title{\bf {Gas molecules (CO, NH$_3$, CO$_2$) sensing with Polyaniline Emardline salt (PANI ES) : a turbo TDDFT study
}}
\date{\today} 
\author{N.~Hadian~Jazi$^{a}$}
\author{I.~Abdolhosseini~Sarsari$^{b,c}$}
\author{N.~Zare~Dehnavi$^{a}$}
\affiliation{a)Department of Physics, Central Tehran Branch, Islamic Azad University, Tehran, Iran\\
b)Department of Physics, Isfahan University of Technology, Isfahan, 84156-83111, Iran\\
c)Computational Physical Sciences Research Laboratory, School of Nano-Science, Institute for Research 
in Fundamental Sciences (IPM), P.O. Box 19395-5531, Tehran, Iran}

\begin{abstract}
The adsorption of gas molecules (CO, NH$_3$, CO$_2$ ) on Polyaniline Emeraldine salt has been performed to study gas sensing.
Density functional theory (DFT) and time dependent density functional theory (TD-DFT) calculations 
have been carried out to compute the response mechanism of polyaniline emeraldine salt (PANI ES)
oligoanalinines (with two to six rings) to gas molecules.
The optimized geometry and electronic structure corresponding to molecule and complexes
are computed with the pseudo-potential and full-potential methods.
The absorption spectra corresponding to polyaniline emeraldine salt molecule and its complexes are
calculated using TD-DFT.
We found it out that the electronic and optical features corresponding to complexes show more sensitive to the NH$_3$ adsorption.
The optical absorption spectrum analysis was used for all ( nPANI ES-X ) and isolated nPANI ES.
Then, the related spectrum indicates that the $\lambda_{max}$ is shifted red or blue, is affected by the kind of complex.

\end{abstract}

\maketitle

\section{INTRODUCTION}

Some gas molecules such as CO, NH$_3$, CO$_2$  can make serious problems, therefore their 
detection at ppm and/or ppb concentration is so important for 
the human safety and health \cite{mahajan1985pollution,timmer2005ammonia,kruse1996calculating,paul2009co,sepaniak2006influence}.
Many articles recently gave emphasis to property of gas sensors.
In the last few years, a new generation of gas
sensors has been prepared, using conductive 
polymers (CP)\cite{bai2007gas,yague2012plasma,penza2004alcohol}.
It is found that the conductive polymers could be utilized for detecting 
small concentration of CO, NH$_3$, CO$_2$ with high sensitivity at ambient temperature \cite{nicolas2003polyaniline}.

Conductive polymers (CPs) (Polyaniline (PANI), polypyrrole (PPY),
poly(o-phenylenediamine) (POPD) and poly(3,4-ethylenedioxythiophene) (PEDOT))
have received a lot of attention since the first
discovery of polyacetylen (CH)$_x$ in 1977\cite{liu2004polymeric,chen2012molecular}.
The many single subject studies demonstrate their biosensing 
and gas sensing capabilities\cite{bai2007gas}.

Among CPs, PANI was studied as gas sensors,
experimentally and theoretically. People hope they will profit greatly from its 
low cost, high sensitivity, environmental stability and short response time
toward guest molecule at ambient temperature in sensing devices\cite{gonccalves2012optical,hoa1992biosensor}.
The response mechanisms of PANI to gas molecules were studied theoretically to give some explanations for empirical observations in detail\cite{virji2006direct,potyrailo2009development,pinto2008electric}.
Recently, more and more theoretical papers have been published
concerning PANI by the mean of Ab-initio or density functional
theory (DFT) methods\cite{choi1999conformational,sein2000theoretical,ullah2013dft,ciric2013recent}.
Geometric and electronic structures of various neutral aniline oligomers
were studied by Lim etal\cite{lim2000comparative}.

\begin{figure}[!ht]
\includegraphics*[scale=1.5]{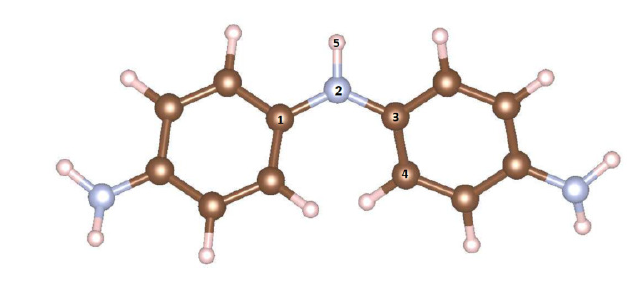}
\caption{\label{2polyaniline}
 Optimized structure of PANI ES molecule. The C, N and H atoms indicate with brown, blue and pink spheres
 respectively. 
}
\end{figure}

In this paper, we studied a variety of complexes to model
the interaction between individual oligoanilines and different gas molecules with (DFT) and Time-Dependent Density Functional Theory (TD-DFT) approaches. 
Based on the synthesis procedure, PANI has three different oxidation states: PANI
Lecuemeraldin base (LB), Pernigraniline base (PNB), Emeradin base (EB).
Acid doping, nitrogen in PANI-EB chain is able to be protonated to afford conducting
Emeraldine salt (ES) which indicates excellent conductivity as the result of extensive $\pi$
conjugating in the polymer chain\cite{macdiarmid2001synthetic}.
In this work, we mainly investigated electronic, optical 
and structural PANI ES, and the effect of interacted gas molecules on it.
Radical cation ES oligoanilines with two, four and six benzene rings ended by nitrogen atoms are represented in Fig. \ref{2polyaniline}.

The general reaction mechanism through which PANI ES gives response to samples X
(X=NH$_3$, CO$_2$, CO) are given based on the subsequent reactions:

\begin{equation}
 \rm PANI ES + x^0 \longrightarrow PANI EB + X^+
\end{equation}
\begin{equation}
 \rm PANI EB + X^+ \longrightarrow  PANI ES + x^0
\end{equation}

\section{Computational Details}

Our electronic structure computation and geometry optimizations 
have been carried out in the frame-work of (DFT) 
by using PAW pseudpopotential as well as numerical orbital atoms (NAO) full potential 
methods applied in the Quantum Espresso (QE)\cite{Giannozzi2009} and FHI-aims\cite{Blum2009}
computational packages, respectively.
GGA-PBE exchange-correlation functionals \cite{perdew1996generalized}
and supercell approach were applied for simulating isolated PANI ES molecule in QE.
A vacuum thickness of around 24 Bohr was used for avoiding interaction
of neighboring molecules. 
Cut-off energy of 40 Ry and 500 Ry, is applied for plane wave development of wave functions
and electron density, However full potential computations were carried out with
$tier3$ base set and atomic ZORA scalar relativistic impacts.

Electronic properties the energy of the highest occupied molecular orbital ( E$_{HOMO}$ ), 
the energy of the lowest unoccupied molecular orbital
( E$_{LUMO}$ ) and HL gap ( E$_{HL}$ ) are calculated at the previously mentioned level of the theory.
The chemical potential ($\mu$) was defined via\cite{rocha2015ab}.
\begin{equation}
 \rm \mu = \frac{-(E_{HOMO} + E_{LUMO})}{2} 
\end{equation}
In addition, the hardness $\eta$ can be calculated by the use of Koopmans’ theorem as:
\begin{equation}
 \rm \eta = \frac{(E_{LUMO} - E_{HOMO})}{2}
\end{equation}
Softness (S) was described using
the following equations\cite{parr1999electrophilicity}:
\begin{equation}
 \rm S = \frac{1}{2\eta}
\end{equation}

To compute the  gas molecules adsorption energy on nPANI ES, Equation (5) has been used. 
\begin{equation}
 \rm E_{ads} = E_{nPANI ES-x} - (E_{nPANI ES} + E_{X}).
\end{equation}
where E$_{nPANI ES-x}$ is the whole energy of the optimized nPANI ES interacted with diverse molecules
of gas, E$_{nPANI ES}$ is the throughout 
energy of an isolated nPANI ES, E$_{X}$ is the full energy of  any gas molecules in the relaxed form.

The optical absorption spectrum of the molecules is calculated in 
the framework of Liouville-Lanczos approach to time-dependent DFT, implemented 
in Turbo-TDDFT code, that is part of QE distribution\cite{malciouglu2011turbotddft,ullrich2014brief}.
The Liouville-Lanczos approach iterations optimized 2000 for our calculations.

\section{RESULTS AND DISCUSIONS}

The interaction of NH$_3$, CO$_2$, CO gas molecules
with PANI ES on different positions have been optimized using the above
mentioned method.\
For NH$_3$ and CO$_2$, just one configuration on PANI ES can
be optimized, However two configurations can be optimized for CO (O end and C end) nPANI ES-OC (CO(1)) and nPANI ES-CO (CO(2)).
Following complete optimization of all configurations on PANI ES, 
the most stable one was applied for further the study.
We find that the divers gas molecules prefer different geometries in the
adsorption. Our data of the optimized geometric parameters and the data of the absorption energy are listed in table\ref{optimized}.
The geometries of the optimized complexes deliver important
data about the response mechanism of conduction
toward different gas molecules.

\begin{table*}[!ht]
\caption{\label{optimized}
Optimized geometric of parameters using DFT level of theory (Bond lengths in \AA{}, bond angles in deg.
}
\begin{ruledtabular}
\begin{tabular}{cccccc}
Species &  $d_{H5..X}$(vdw)  & $d_{N2..H5}$  & $\angle$C$_{1}$N$_{2}$C$_{3}$ & $\angle$N$_{2}$H$_{5}$..X  & $\angle$C$_{1}$N$_{2}$C$_{3}$C$_{4}$   \\
\hline
Isolated 2PANI ES & -    & 1.01 & 129.04 &   -    & 22.27\\
2PANI ES-CO(2)    & 2.19 & 1.01 &  129.01 & 179.75 & 22.29  \\
2PANI ES-CO(1)    & 2.12 & 1.02 &  128.58 & 179.91 & 23.07 \\
2PANI ES-CO$_{2}$ & 2.07 & 1.01 &  129.16 & 179.98 & 21.86 \\
2PANI ES-NH$_{3}$ & 1.87 & 1.04 &  127.71 & 179.29 & 24.49 \\
Isolated 4PANI ES & -    & 1.01 &  130.36 &    -   & 22.58  \\
4PANI ES-CO(2)    & 2.12 & 1.02 & 129.02  & 179.25 & 22.67   \\
4PANI ES-CO(1)    & 2.12 & 1.01 & 129.04  & 179.15 & 22.35   \\
4PANI ES-CO$_{2}$ & 2.00 & 1.01 &  129.57 & 179.78 & 20.97   \\
4PANI ES-NH$_{3}$ & 1.84 & 1.04 & 128.29  & 178.66 & 23.04   \\
Isolated 6PANI ES & -    & 1.01 &  129.53 &   -    & 21.65   \\
6PANI ES-CO(2)    & 2.08 & 1.01 &  129.50 & 179.82 & 20.78   \\
6PANI ES-CO(1)    & 2.13 & 1.02 &  128.95 & 179.78 & 22.13  \\
6PANI ES-CO$_{2}$ & 1.99 & 1.01 &  129.53 & 179.68 & 21.44  \\
6PANI ES-NH$_{3}$ & 1.84 & 1.04 &  128.43 & 177.83 & 23.28  \\
\end{tabular}
\end{ruledtabular}
\end{table*}

The distance between H5 and X gas ($d_{H5..X}$) shows notable changes in different gas molecules adsorption. The nearest distance between nPANI ES and gas molecules is confirmed for NH$_{3}$ by the distance of 1.87\AA{}, 1.84\AA{}, 1.84 \AA{} for n=2,4,6, respectively, which points to strong interaction between them but CO(1) and CO(2) have the largest distance which points to less interaction between them. 
This result is completely in agreement with the obtained adsorption energies.
For nPANI ES-NH$_{3}$ complexes, $d_{H5..X}$ distance decreases about 0.03\AA{} from 2 to 6. The reduction of $d_{H5..X}$ distance with chain length elongation illustrates the reduction in interaction with chain elongation.
The distance $d_{N2..H5}$ increases when nPANI ES interacts with ammonia (table \ref{optimized}). This bond enhance/reduce as the result of generating/losing of ion dipole electrostatic interaction between nPANI ES and the above-mentioned molecules.

Another important geometric factor in nPANI ES and nPANI ES-X is the bridging angle ($\angle$C$_{1}$N$_{2}$C$_{3}$)  which decrease when nPANI ES interacts with NH$_{3}$ molecule. The maximum and minimum decrease in angle is about 2.07 \AA{} and 1.1 \AA{} in 4PANI ES-NH$_{3}$ and 6PANI ES-CO(1) complexes regarding their equivalent non-complexed nPANI ES. The larger decrease in angle means the stronger electrostatic attraction between nPANI ES and NH$_{3}$. 
considering the analysis of the decrease in bond angles, it can be concluded that the interaction between nPANI ES and NH$_{3}$ increase and alongside the increase in chain length elongation, consistent with the inference from the bond length analysis (vide supra). 

Another important parameter is the $\angle$N$_{2}$H$_{5}$..X angle which only 
shows the situation of gas molecules on nPANI ES. it increases from about 129 degree 
(in nPANI ES) to about 180 degree in all nPANI ES-X complexes. 
As the result of existing different gas molecules which is presented
in table \ref{optimized} the dihedral angle $\angle$C$_{1}$N$_{2}$C$_{3}$C$_{4}$ is altered based on the kind of X as guests which are posiibly beacuse of the ion dipole interaction. 
This factor will influence the electronic structure of complexes.

\subsection{Electronic properties}

The electronic structure corresponding to the best nPANI ES molecule and its complexes were computed 
with both pseudo-potential and full potential methods and result showed an excellent was observed.
The electronic properties of nPANI ES are completely modified by gas molecules.
The prime focus of the present work is on ionization electron, electron affinity, 
HOMO-LUMO(HL) gap and adsorption characteristic of gas molecules 
on nPANI ES. Table \ref{geometry} represents the HOMO energy, LUMO energy, 
Dipole moment, Mulliken charge transfer, HL gap and adsorption 
energy of isolated nPANI ES and nPANI ES complexes. The HL gap  
of isolated nPANI ES is 2.71, 2.31 and 2.36 eV for n=2, 4, 6. 

The incorporation of gas molecules decreases the E$_{HL}$  
except for 4PANI ES-NH$_{3}$. So we conclude that for n =2-6 the 
highest conductivity between nPANI ES and different gas molecules
are confirmed for CO(1), CO(2). Analyzing the HL gap variation upon 
gas molecules adsorption on nPANI ES isolated clearly suggests that 
nPANI ES can be used as a sensor. As can be observed in table \ref{geometry}, 
the HL gap of each gas molecules is much higher than the HL gap of nPANI ES.

However, upon complexation, the HL gap of the resultant complex is
quite different from that one for of the nPANI ES, depending on
the nature of the gas molecules. 
It can be said (but it is not true for all cases) that the smaller difference in the HL gap of
gas molecules and nPANI ES molecule, the greater change in HL gap
of their complex (further hybridization), and as it can be observed
for CO(2), CO(1). For other samples like CO$_{2}$ the large
difference in the E$_{HL}$ of the gas molecule and nPANI ES results in an
inconsiderable variation in the HL gap of the resultant complex, which makes it difficult to transfer electrons\cite{chattaraj2009electrophilicity}.

When gas molecules get adsorbed on nPANI ES, the alteration in adsorption energy would be observed. 
The positive and negative values of $E_{ad}$ show the endothermic and exothermic reaction
respectively, based on the adsorption of gas molecules on nPANI ES. 
The adsorption energy magnitude is different and a negative value of adsorption 
energy infers that the energy transfers from isolated nPANI ES 
to gas molecules.
The adsorption energy magnitude for 
nPANI ES-NH$_{3}$ is higher than others.

The electronic features can also be discussed in terms of Mulliken charge transfer (Q) between gas molecules
and isolated nPANI ES. The positive value corresponding to Mulliken charge transfer indicates the charge tansfer from gas  molecules to nPANI ES, while the negative value corresponding to Mulliken charge transfer indicates the charge transfer from nPANI ES isolated to gas molecules.
In the present work, for all positions, the positive value of Mulliken charge transfer is taken in to consideration and indicates the electrons transfer from gas molecules to nPANI ES. 
The magnitude of Mulliken charge transfer is found to be higher when 
NH$_{3}$ molecule get adsorbed on nPANI ES  comparing to 
the adsorption of other molecules and it is due to more value of 
adsorption energy for nPANI ES-NH$_{3}$ that can be observed in table \ref{geometry}.

A weak interaction is observed in nPANI ES-CO$_{2}$ complexes,
comparing those of nPANI ES-NH$_{3}$ (see table \ref{geometry}). The
HOMO of 2PANI ES gets electric charge clouds from CO$_{2}$
of nearby 0.13 eV which is 0.2 eV smaller comparing to that of the 2PANI ES-NH$_{3}$ complex. 
The transfer of this electron clouds happen from the HOMO of CO$_{2}$ to LUMO of nPANI ES, instead of ion-dipole electrostatic interaction with ammonia,  give a rise to a weak type of attractive force.
The small electron cloud
transfer from analyte to polymer in 2PANI ES-CO$_{2}$ comparing
to 2PANI ES-NH$_{3}$ complex manifests itself in weak sensing of CO$_{2}$ using the polymer.
The 4PANI ES-CO$_{2}$ and 6PANI ES-CO$_{2}$ complexes
have 0.17 eV and 0.03 eV smaller electron cloud attraction,
respectively, as comparing to its nPANI ES-NH$_{3}$.

\begin{table*}[!ht]
\caption{\label{geometry}
The calculated  electronic properties of of nPANI ES and nPANI ES-X complexes, where n = 2-6 and
X = NH$_{3}$, CO(1), CO(2) and CO$_{2}$,
$E_{g}$ = $\lvert E_{HOMO} - E_{LUMO}\rvert$, $E_{d}$: adsorption energy, $Q(e)$: Mulliken charge transfer,
$\mu_{D}$: dipole moment.
}
\begin{ruledtabular}
\begin{tabular}{ccccccc}
Species & $E_{HOMO}$(eV) & $E_{LUMO}$(eV) & $E_{HL}(eV)$ & $E_{ad}(eV)$ & $Q(e)$ & $\mu_{D}(Debye)$\\
\hline
Isolated 2PANI ES & -3.69 & -0.97 &  2.72 &   -    &     -   & -0.22 \\
2PANI ES-CO(2)    & -3.63 & -2.50 &  1.12 & -0.08 & 0.02   & -0.39\\
2PANI ES-CO(1)    & -3.60 & -2.59 &  1.00 & -0.16 & 0.03   & -0.60\\
2PANI ES-CO$_{2}$ & -3.56 & -1.15 &  2.41 & -0.12 & 0.01   & -0.90\\
2PANI ES-NH$_{3}$ & -3.36  & -0.95 &  2.41 & -0.41 & 0.06   & -3.00\\
Isolated 4PANI ES & -3.39 & -1.07 &  2.31 &   -    &  -      & 24.76\\
4PANI ES-CO(2)    & -3.41 & -2.47 &  0.93 & -0.17 & 0.03   & 27.22\\
4PANI ES-CO(1)    & -3.42 & -2.39  &  1.03 & -0.08 & 0.02   & 27.82\\
4PANI ES-CO$_{2}$ & -3.37 & -1.06 &  2.31 & -0.13 & 0.01   & 31.80\\
4PANI ES-NH$_{3}$ & -3.23 & -0.89 &  2.34 & -0.43 & 0.06   & 32.67\\
Isolated 6PANI ES & -3.35 & -0.98 &  2.36 &   -    &     -   & -1.11\\
6PANI ES-CO(2)    & -3.37 & -2.41 &  0.95 & -0.08 & 0.02   & 23.59\\
6PANI ES-CO(1)    & -3.36 & -2.49 &  0.87 & -0.17 & 0.03   & 22.04\\
6PANI ES-CO$_{2}$ & -3.33 & -1.07 &  2.25 & -0.13 & 0.02   & 22.59\\
6PANI ES-NH$_{3}$ & -3.22 & -0.97 &  2.25 & -0.50 & 0.06   & 25.29\\
CO(2)             & -9.11 & -2.32  &  6.78 &   -    &   -     & 0.16 \\
CO(1)             & -9.10 & -2.31 &  6.79 &   -    &   -     & -0.17 \\
CO$_{2}$          & -9.34  & -1.01 &  8.32 &   -    &   -     & -0.01 \\
NH$_{3}$          & -6.12 & -0.74   &  5.38 &   -    &   -     & -1.37 \\
\end{tabular}
\end{ruledtabular}
\end{table*}

The computed amounts of dipole moment $\mu_{D}$ of interacted 
CO(1), CO(2), CO$_{2}$ and NH$_{3}$ with nPANI ES are presented in 
table \ref{geometry}\cite{rad2016terthiophene}. 
The dipole moment considers the electronic and geometrical features and 
gives an insight into the distribution of charges in alone Polyaniline emardline salt. 
Our findings indicated that upon interaction of analytes with nPANI ES
the dimension and directions of $\mu_{D}$  will be changed based on the configuration of adsorption. 
As presented in table \ref{geometry}, for each system of interaction with nPANI ES, the dipole moment 
increased compare to isolated nPANI ES.

Orbital density of nPANI ES molecule and several of its derivatives are computed and 
showed in table \ref{HL}. The main activity of electrons is in the state of HOMO. 
Therefor we are pretty zealous to study the HOMO-LUMO orbitals of all structures. 
As it is presented in table \ref{HL}.
The distribution of HOMO and LUMO (See table \ref{HL})
for nPANI ES and complexed forms of species are: nPANI ES-CO(2), 
nPANI ES-CO(1), nPANI ES-CO$_{2}$ and nPANI ES-NH$_{3}$ for (n=2,4,6). 
As can be seen in table \ref{HL} for all systems, HOMO
is localized on nPANI ES molecule except for 4PANI ES-CO(2) and 
6PANI ES-CO(1), the HOMO are located on both CO(2), CO(1) and 
4PANI ES, 6PANI ES-CO(1). 
This relocation of HOMO corresponds to the highest change in the E$_{HL}$ and causes the
highest interaction energy. Similarly, for 2PANI ES, 4PANI ES and 6PANI ES, 4 PANI ES-CO$_{2}$ 
the LUMO is located on nPANI ES molecule due to weak interaction between nPANI ES with gas molecules.
The LUMO for 2PANI ES-CO(2), 2PANI ES-CO(1), 4PANI ES-CO(2), 4PANI ES-CO(1),6PANI ES-CO(2) and 6PANI ES-CO(1) are throughly located on the CO(2), CO(1) with an exception for 2PANI ES-CO$_{2}$, 2PANI ES-NH$_{3}$, 6PANI ES-CO$_{2}$, 4PANI ES-NH$_{3}$, the LUMO  which are localized both on CO$_{2}$, NH$_{3}$ and polymer rings. This relocation of LUMO corresponds to the highest change in the E$_{HL}$, that results in the
highest interaction energy.

\begin{table*}[!ht]
\caption{\label{HL} 
Orbital density of nPANI ES molecule and its derivatives.
}
\begin{ruledtabular}
\begin{tabular}{cccccc}
Phase  &  Isolated 2PANI ES                               & 2PANI ES-CO(2)                & 2PANI ES-CO(1)     & 2PANI ES-CO$_{2}$  & 2PANI ES-NH3$_{3}$ \\
\hline
LUMO   & \includegraphics*[scale=0.06]{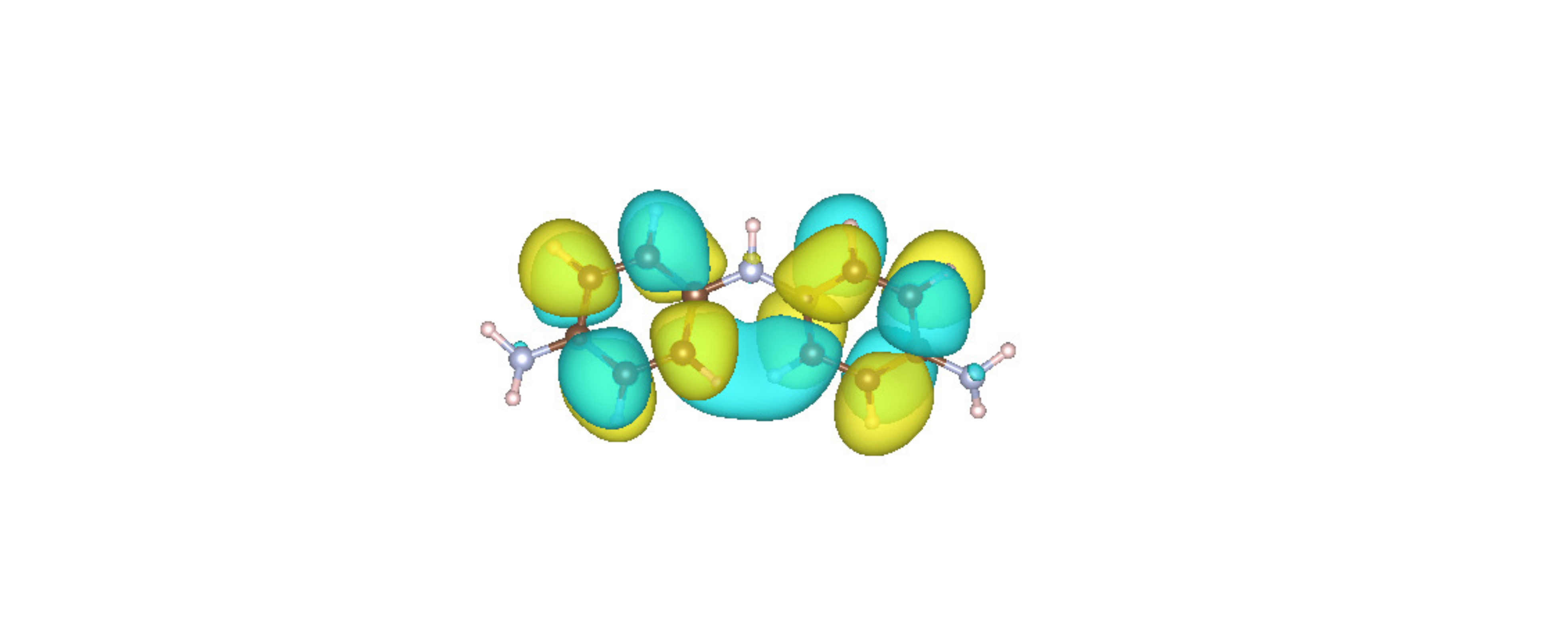} & \includegraphics*[scale=0.06]{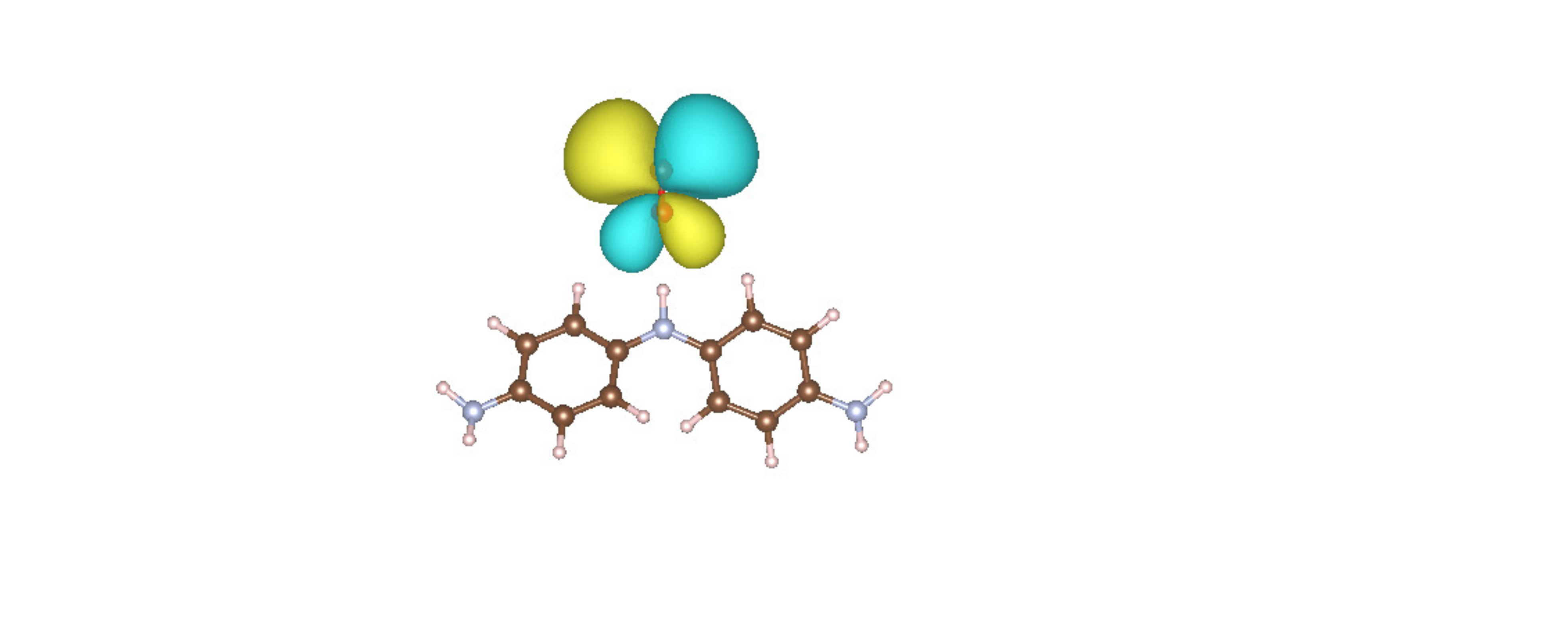} & \includegraphics*[scale=0.06]{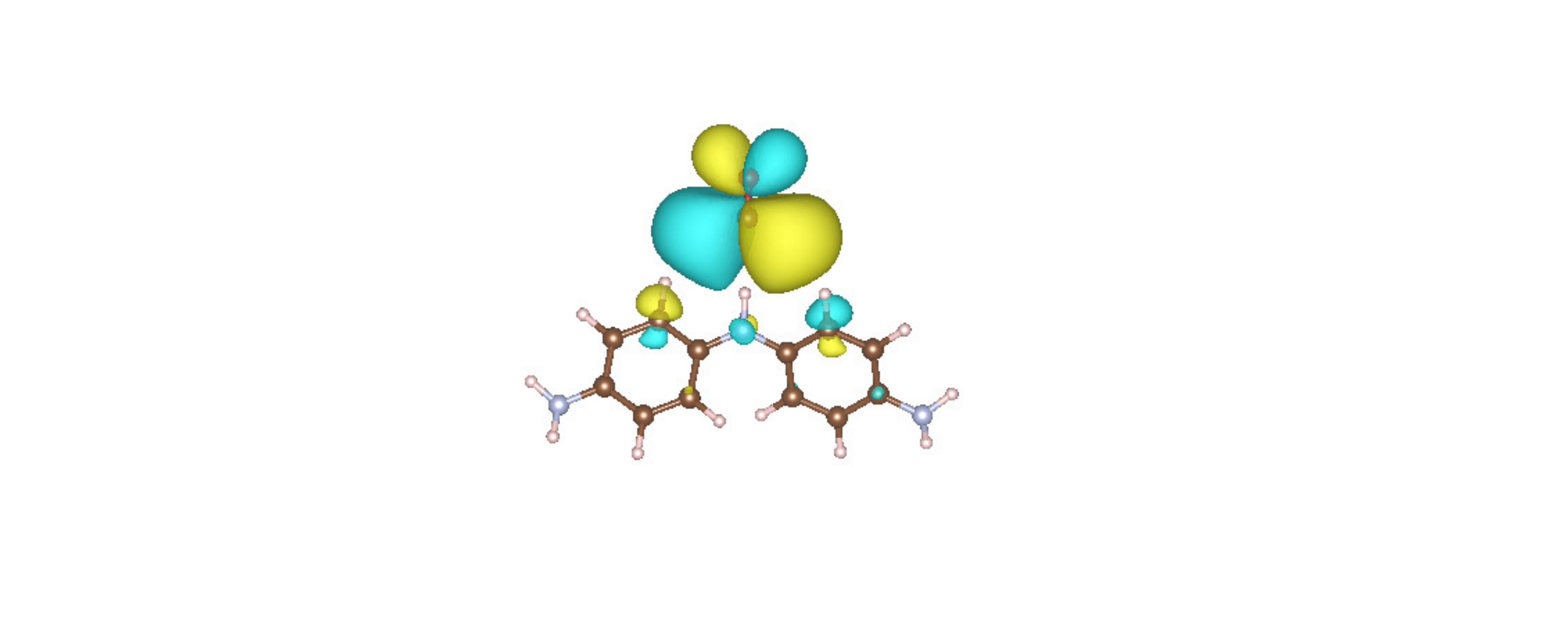} & \includegraphics*[scale=0.06]{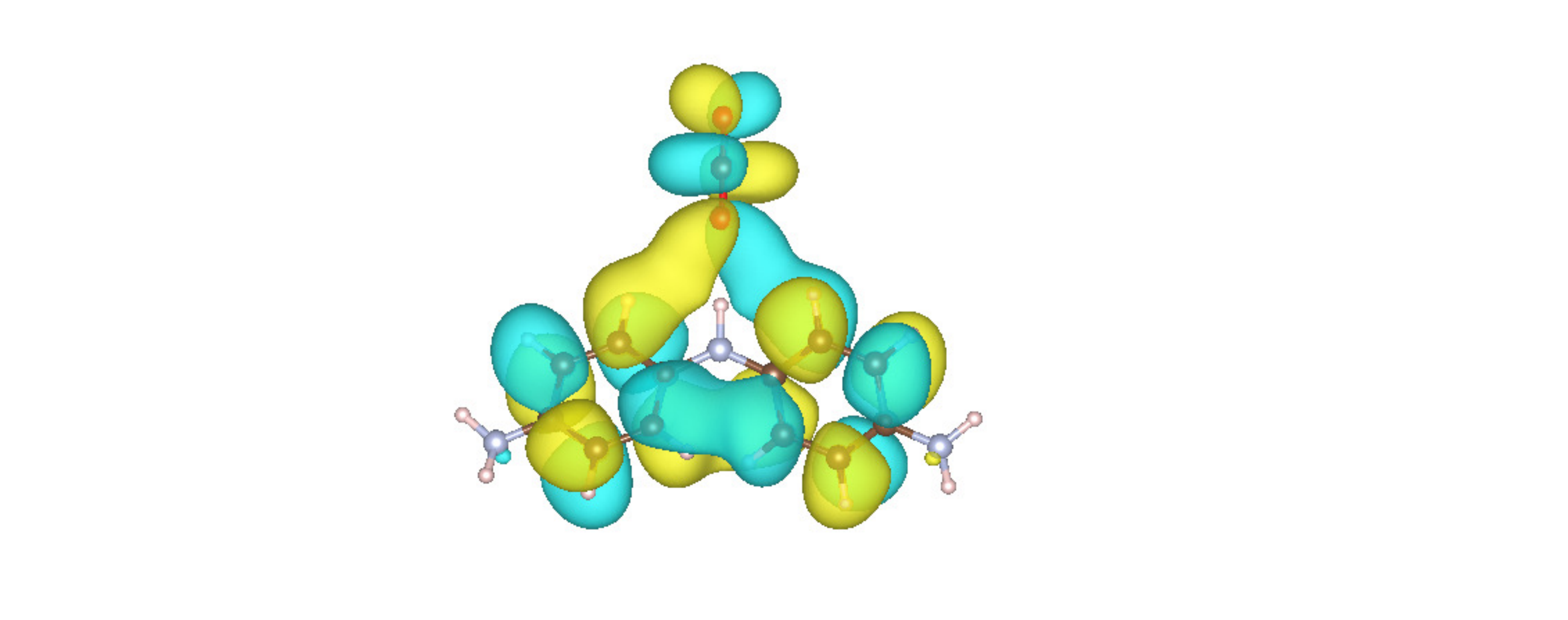} & \includegraphics*[scale=0.06]{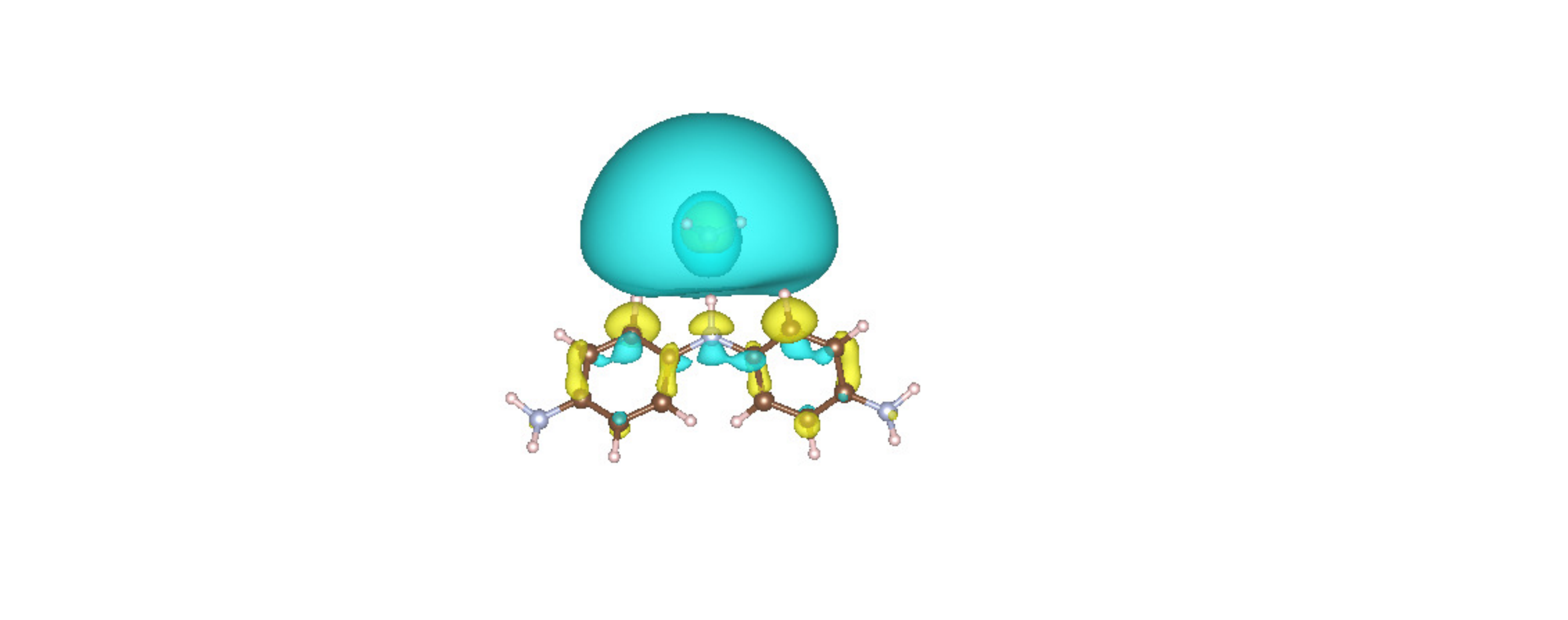} \\
HUMO   & \includegraphics*[scale=0.06]{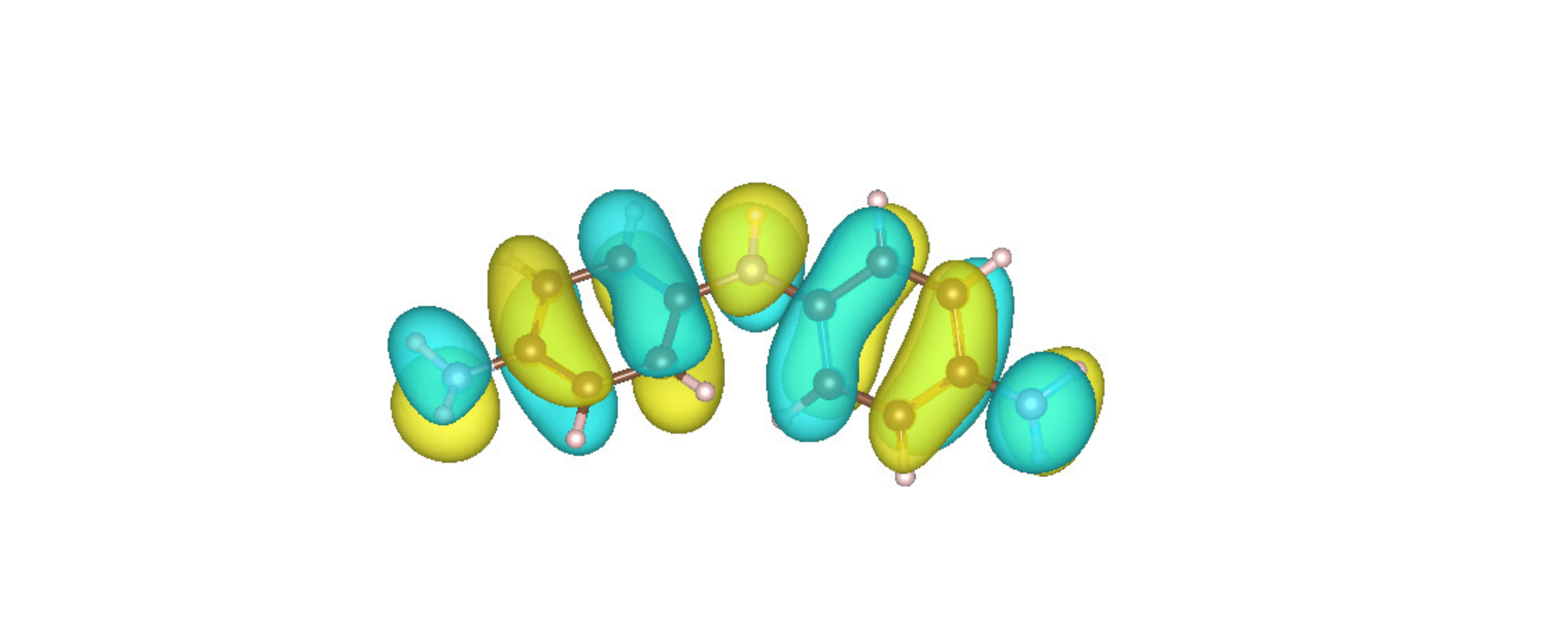} & \includegraphics*[scale=0.06]{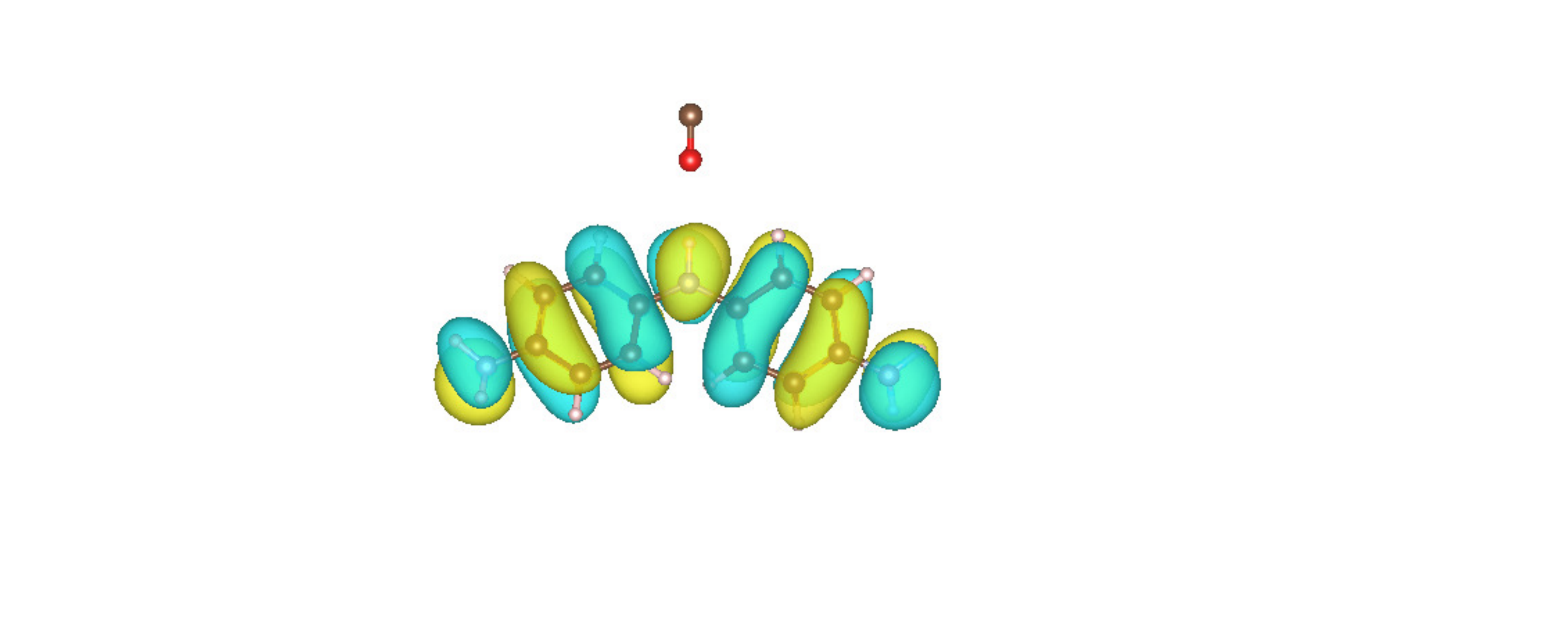} & \includegraphics*[scale=0.06]{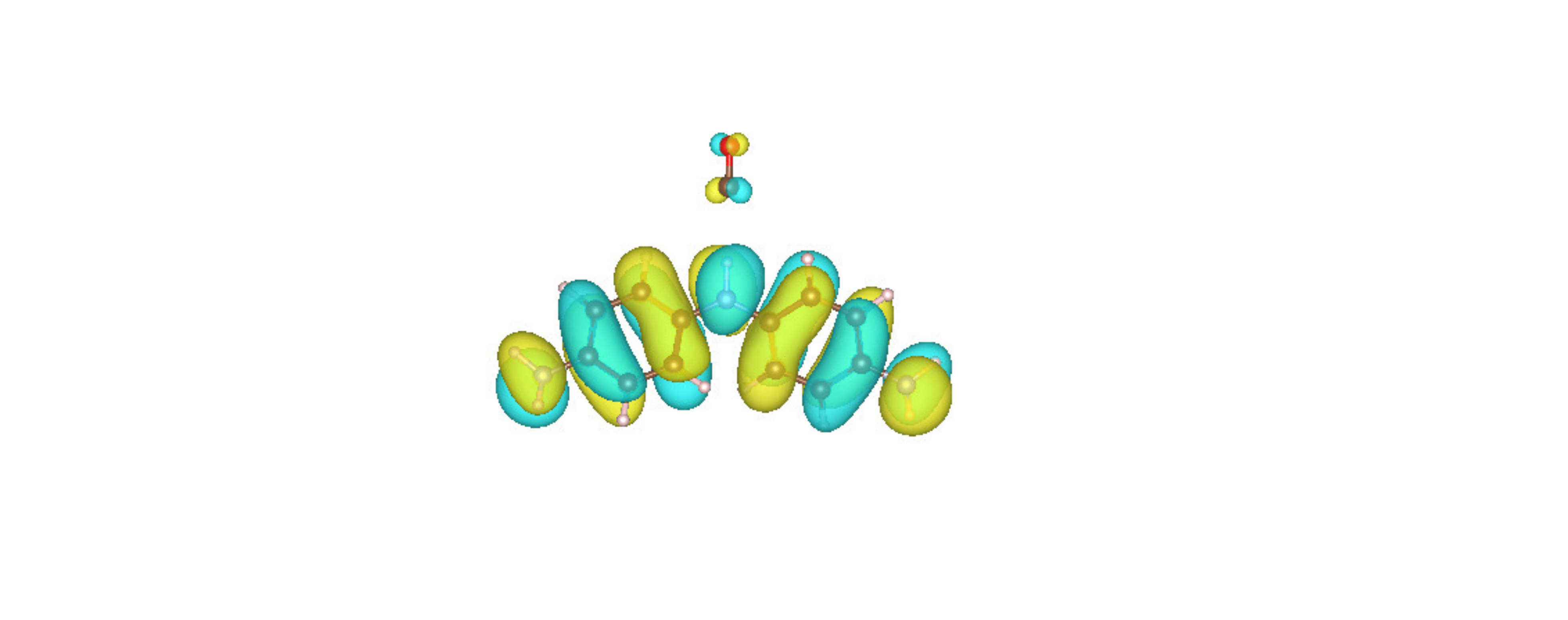} & \includegraphics*[scale=0.06]{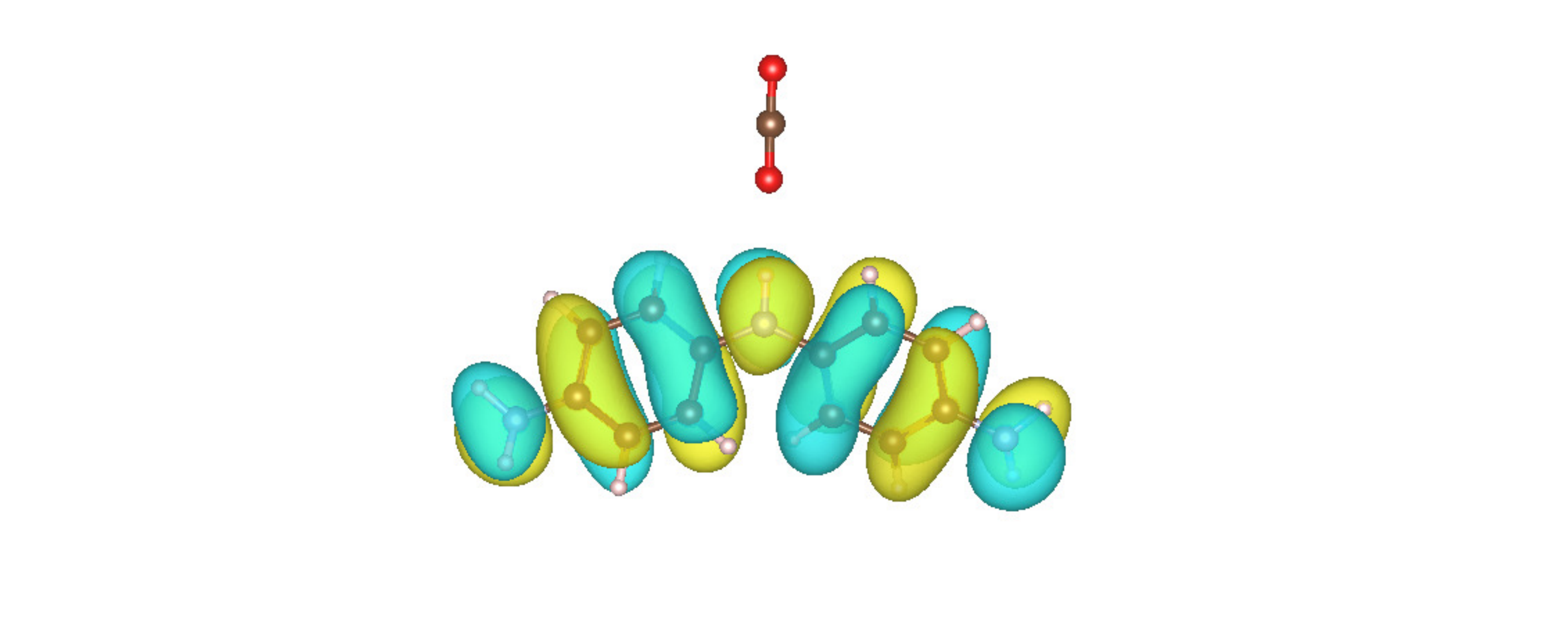} & \includegraphics*[scale=0.06]{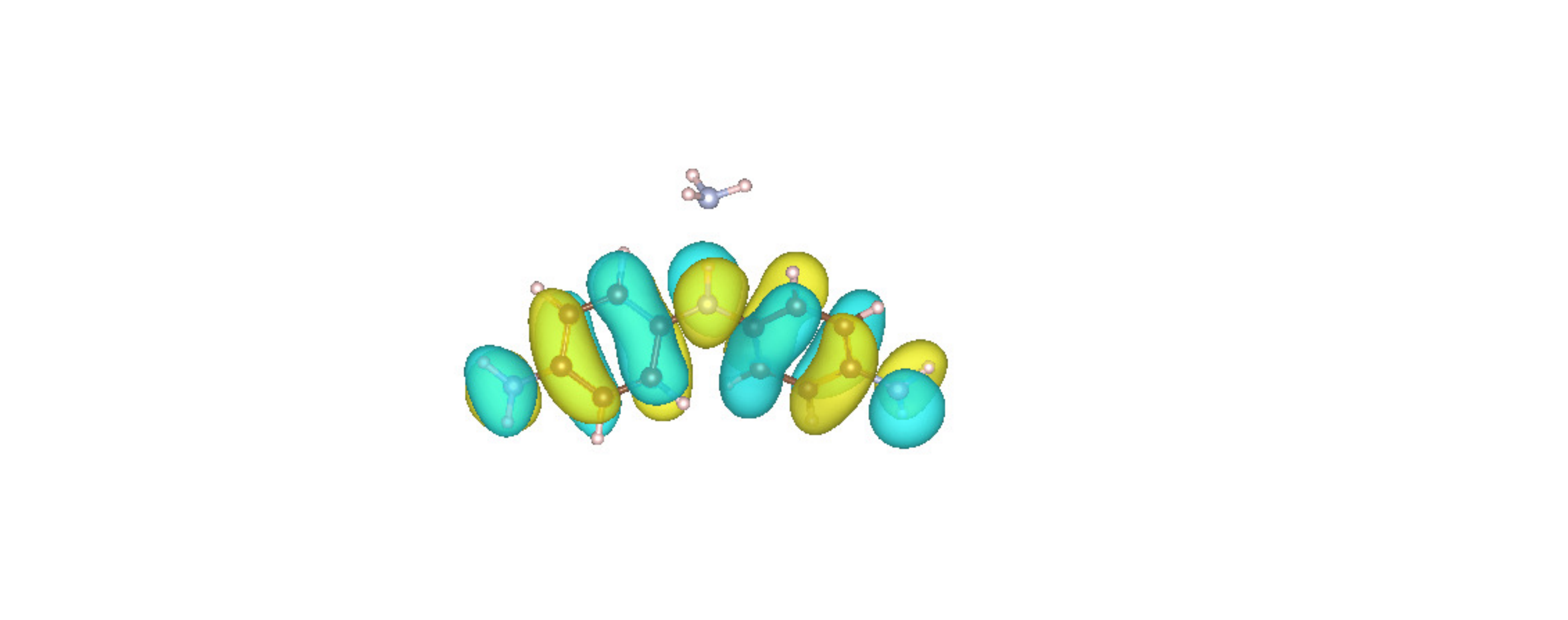}  \\                                  
\hline
\hline
Phase      & Isolated 4PANI ES         & 4PANI ES-CO(2)       & 4PANI ES-CO(1)     & 4PANI ES-CO$_{2}$   & 4PANI ES-NH3$_{3}$    \\
\hline
LUMO       &  \includegraphics*[scale=0.06]{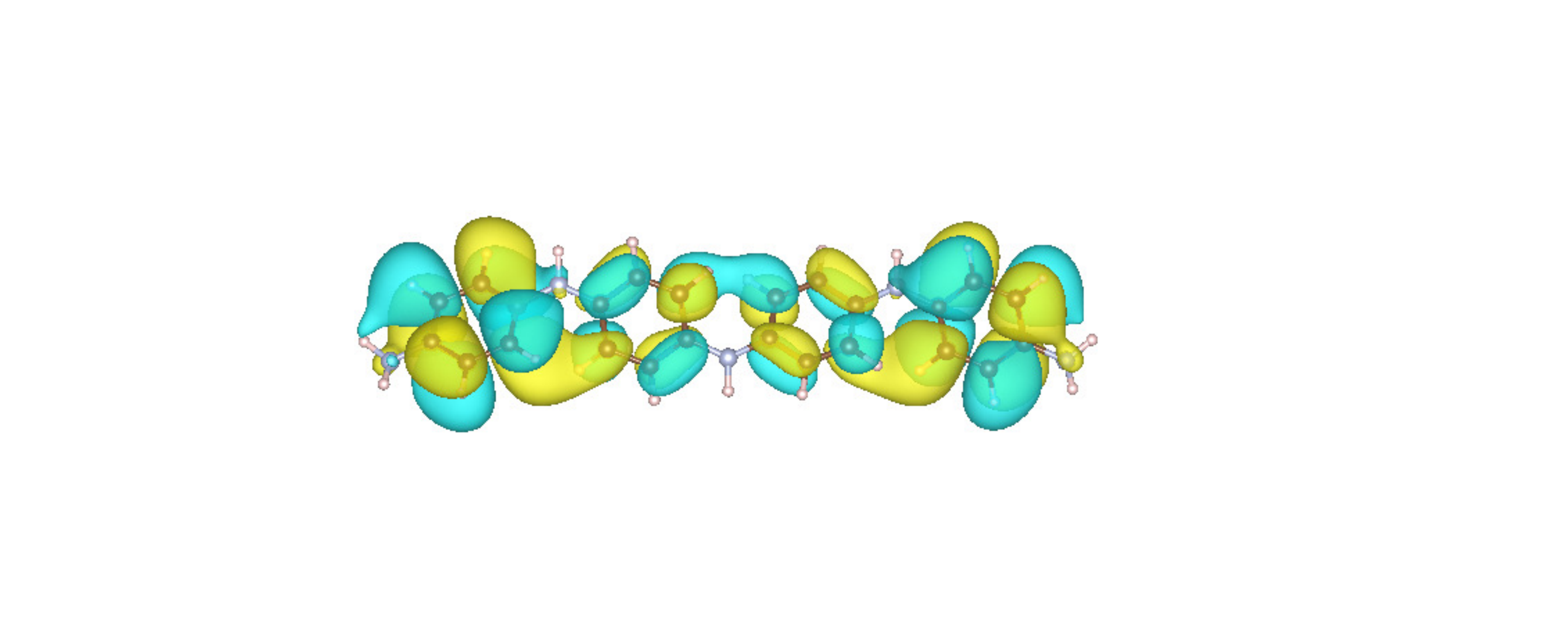} & \includegraphics*[scale=0.06]{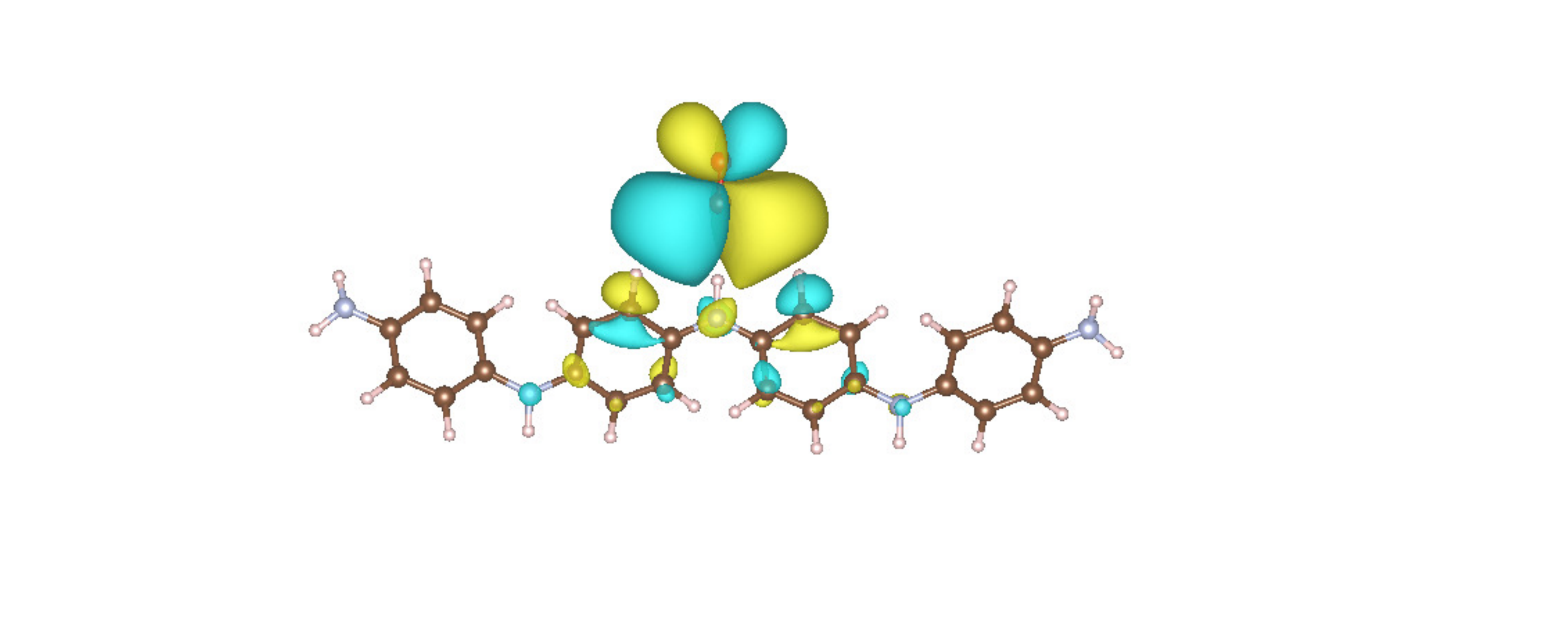}  & \includegraphics*[scale=0.06]{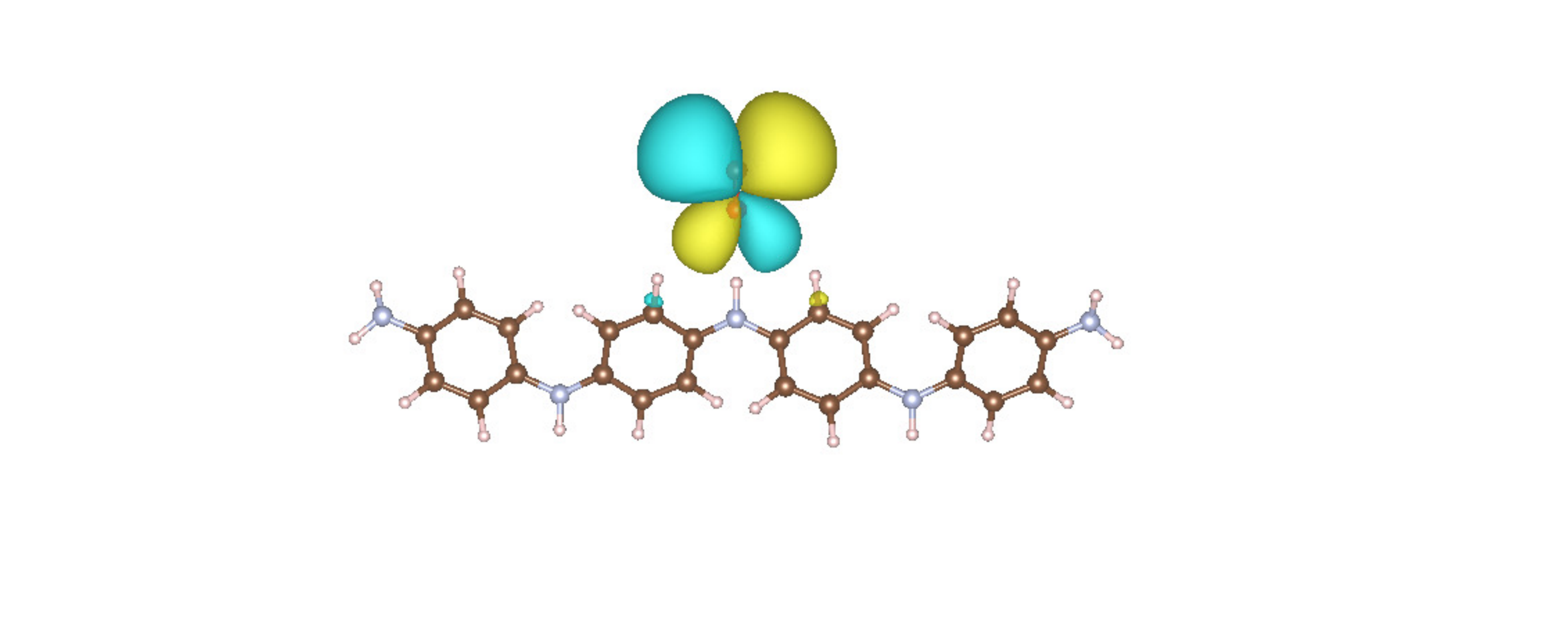} & \includegraphics*[scale=0.06]{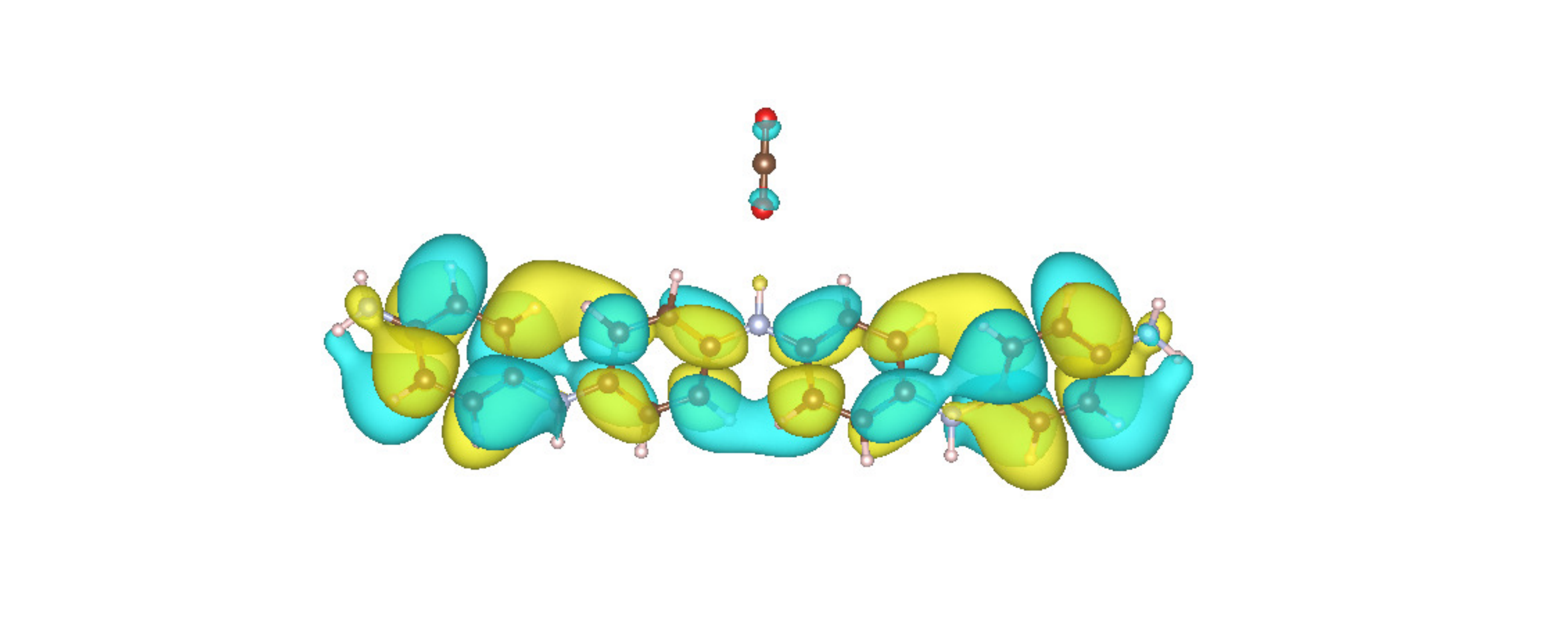} & \includegraphics*[scale=0.06]{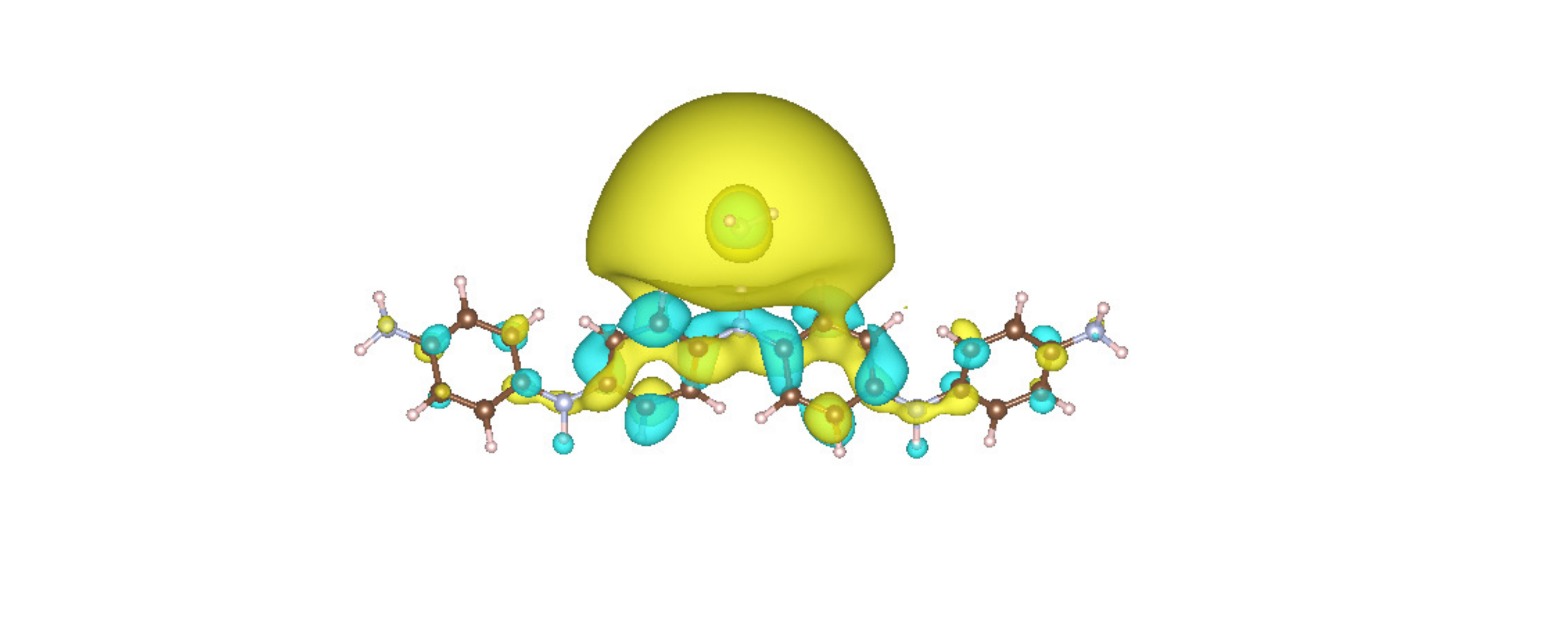} \\
HUMO       &  \includegraphics*[scale=0.06]{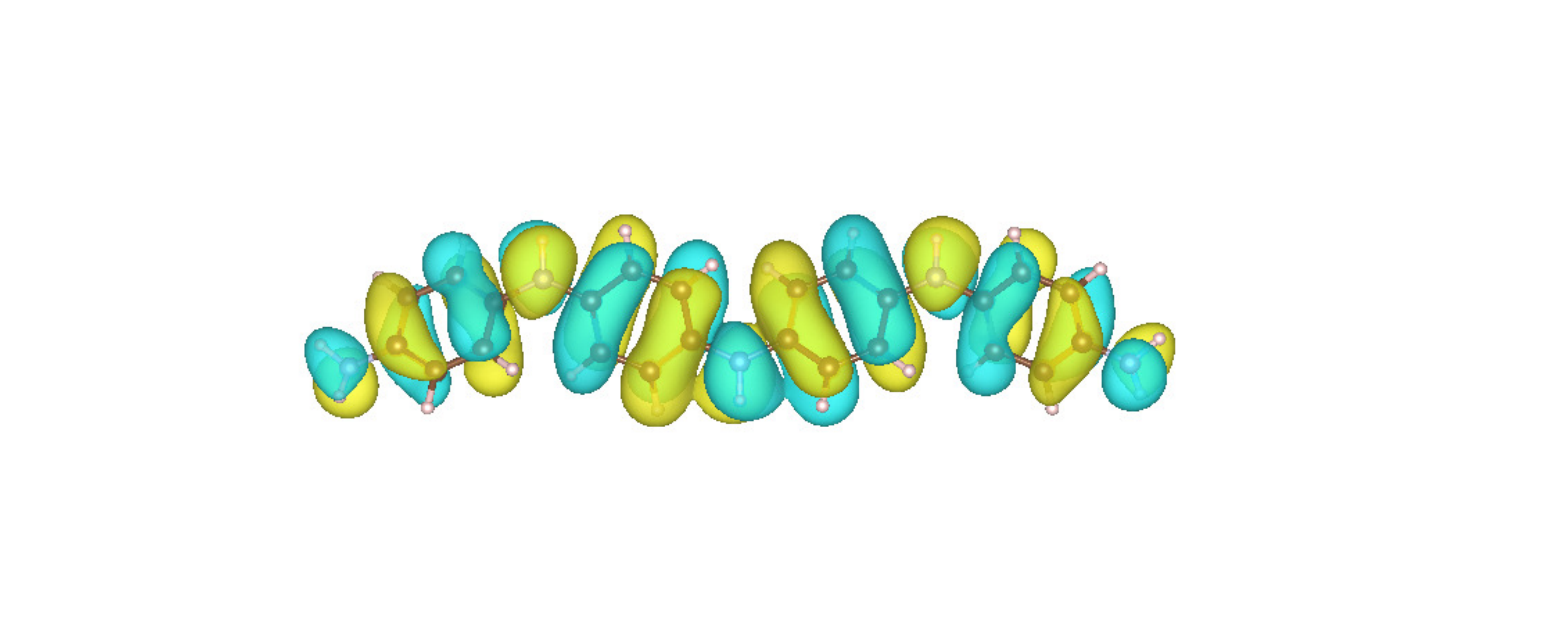} & \includegraphics*[scale=0.06]{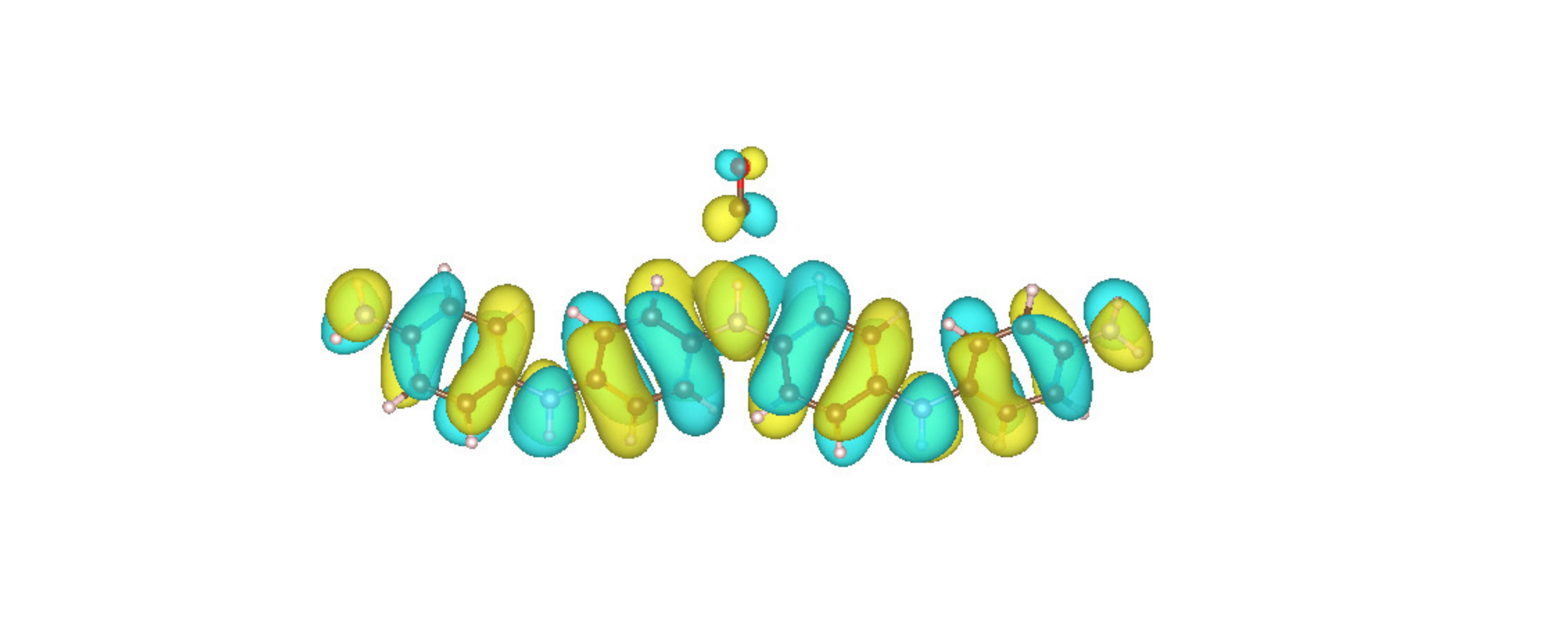}  & \includegraphics*[scale=0.06]{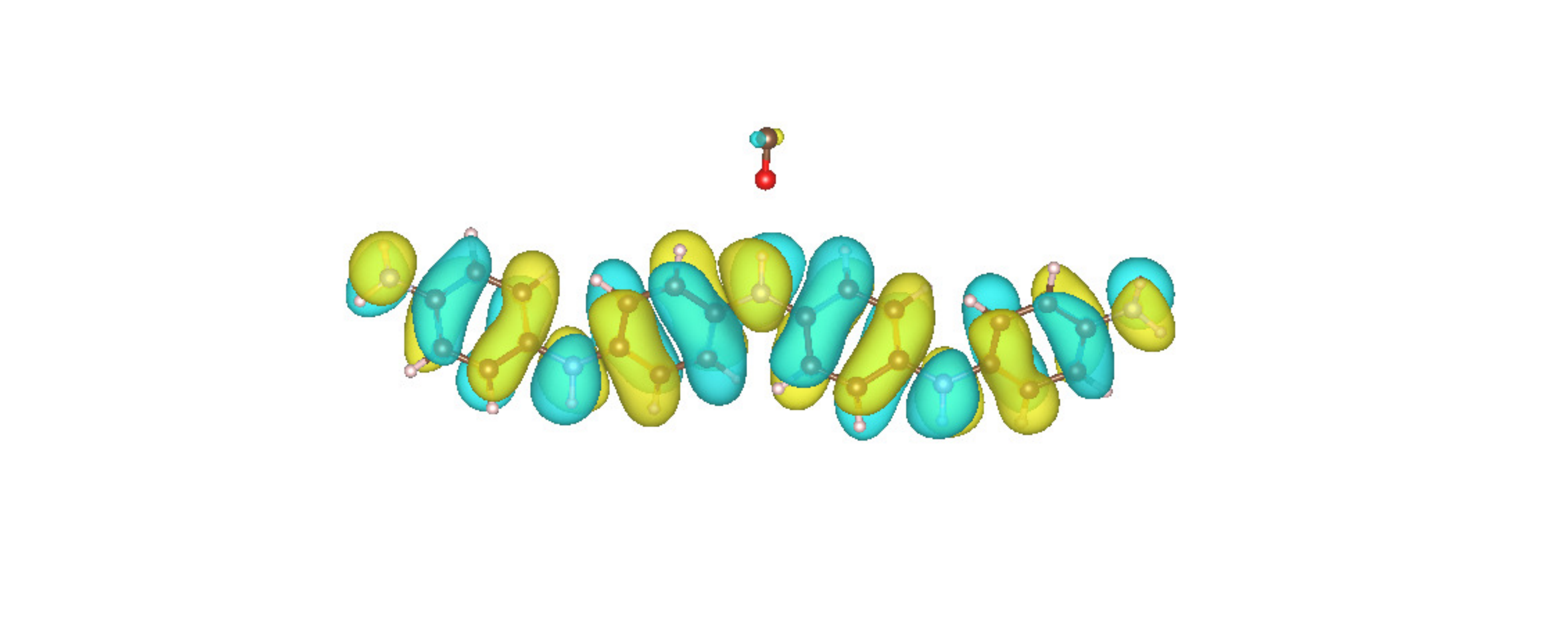} & \includegraphics*[scale=0.06]{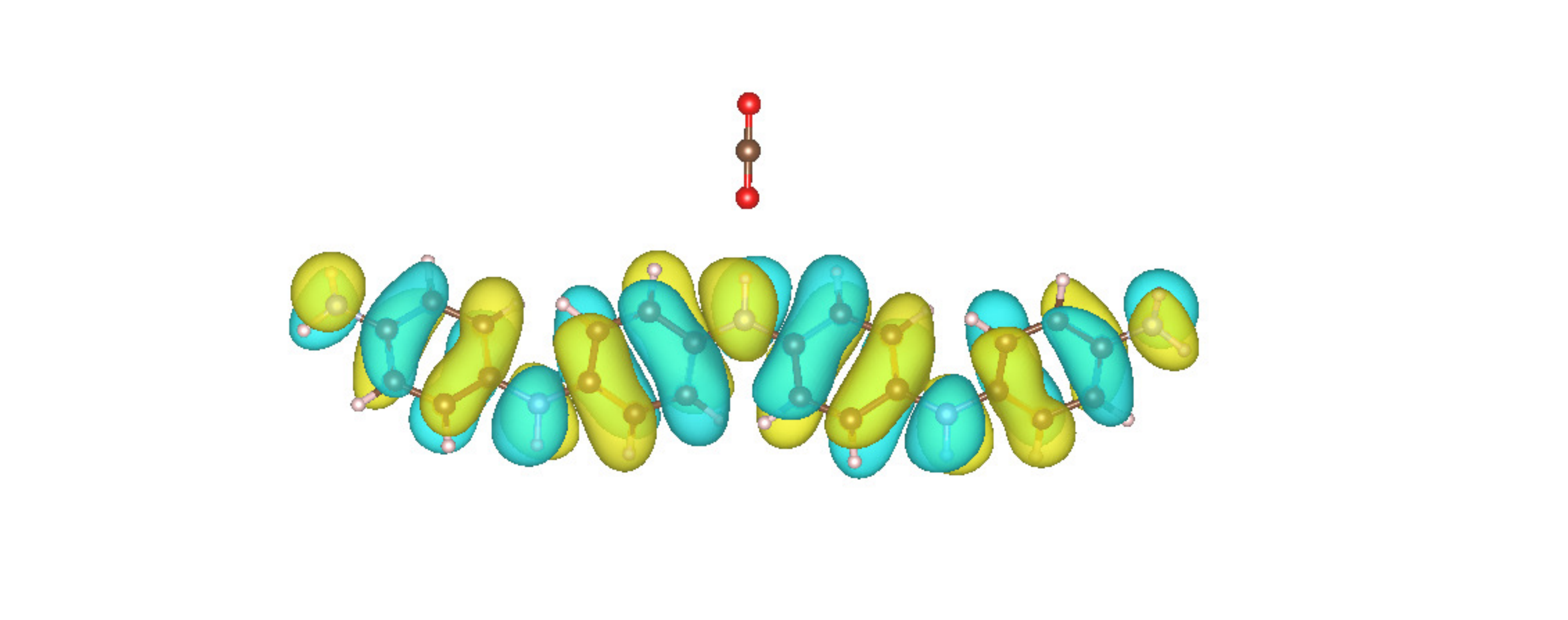} & \includegraphics*[scale=0.06]{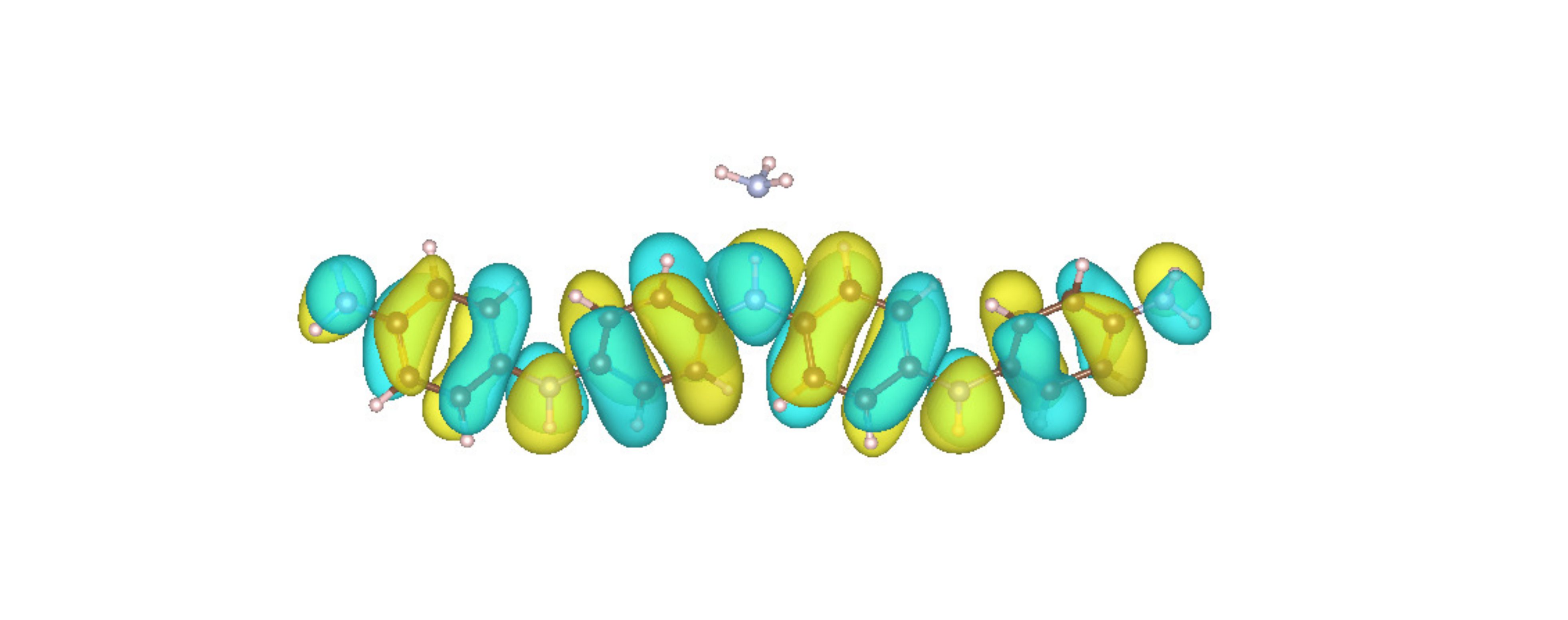} \\                                          
\hline
\hline
Phase      & Isolated  6PANI ES         & 6PANI ES-CO(2)       & 6PANI ES-CO(1)     & 6PANI ES-CO$_{2}$   & 6PANI ES-NH3$_{3}$    \\
\hline
LUMO       &  \includegraphics*[scale=0.06]{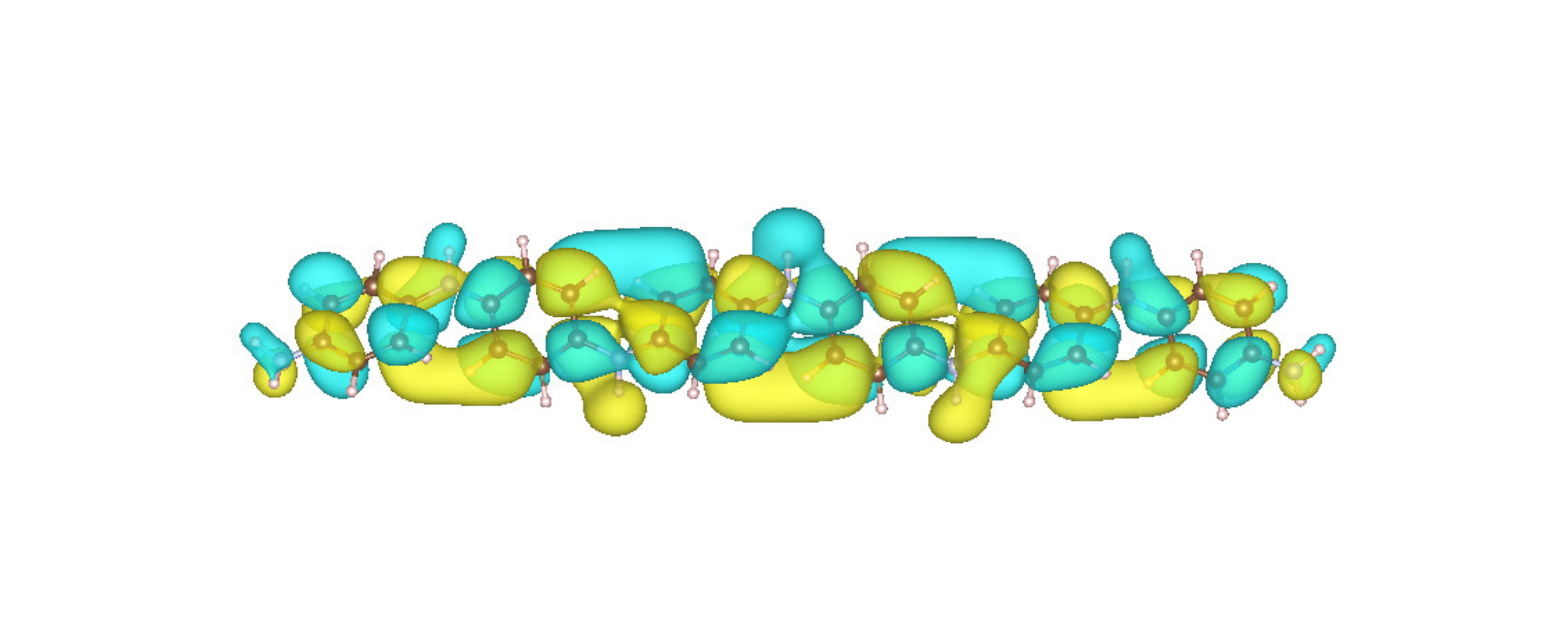} & \includegraphics*[scale=0.06]{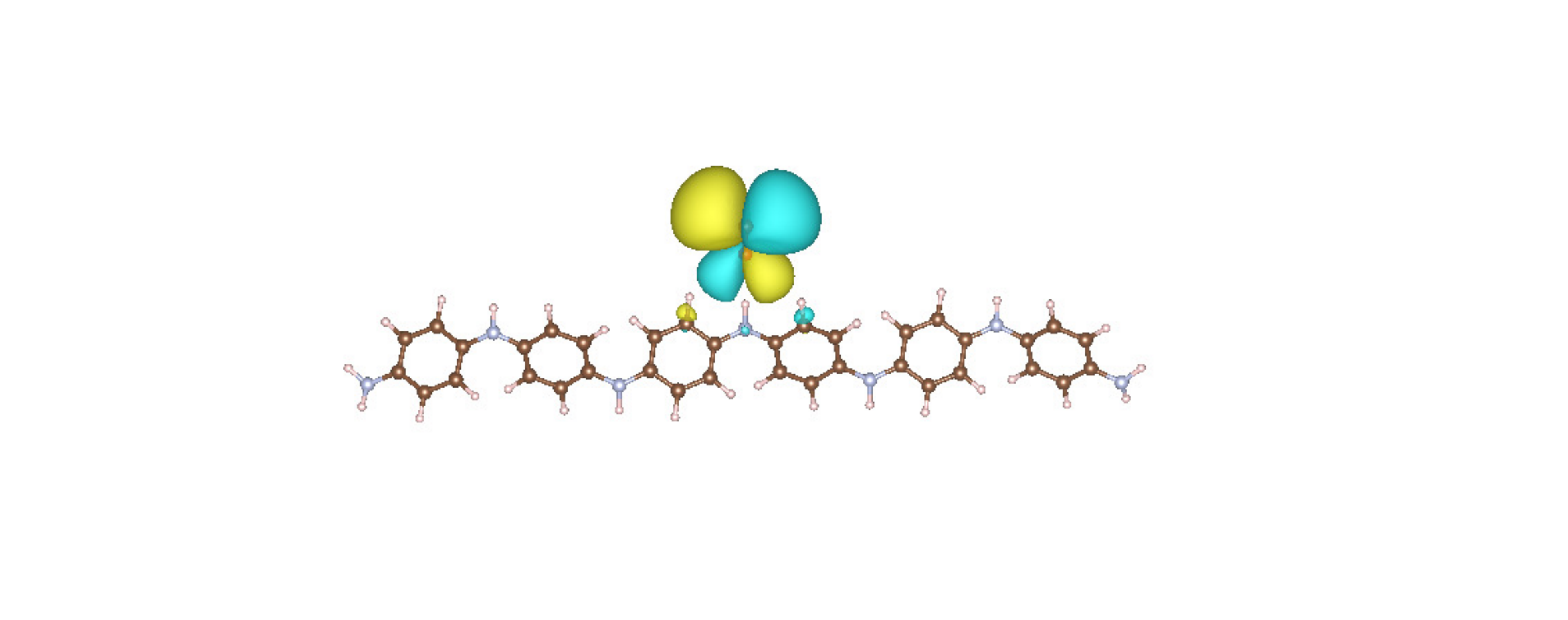}  & \includegraphics*[scale=0.06]{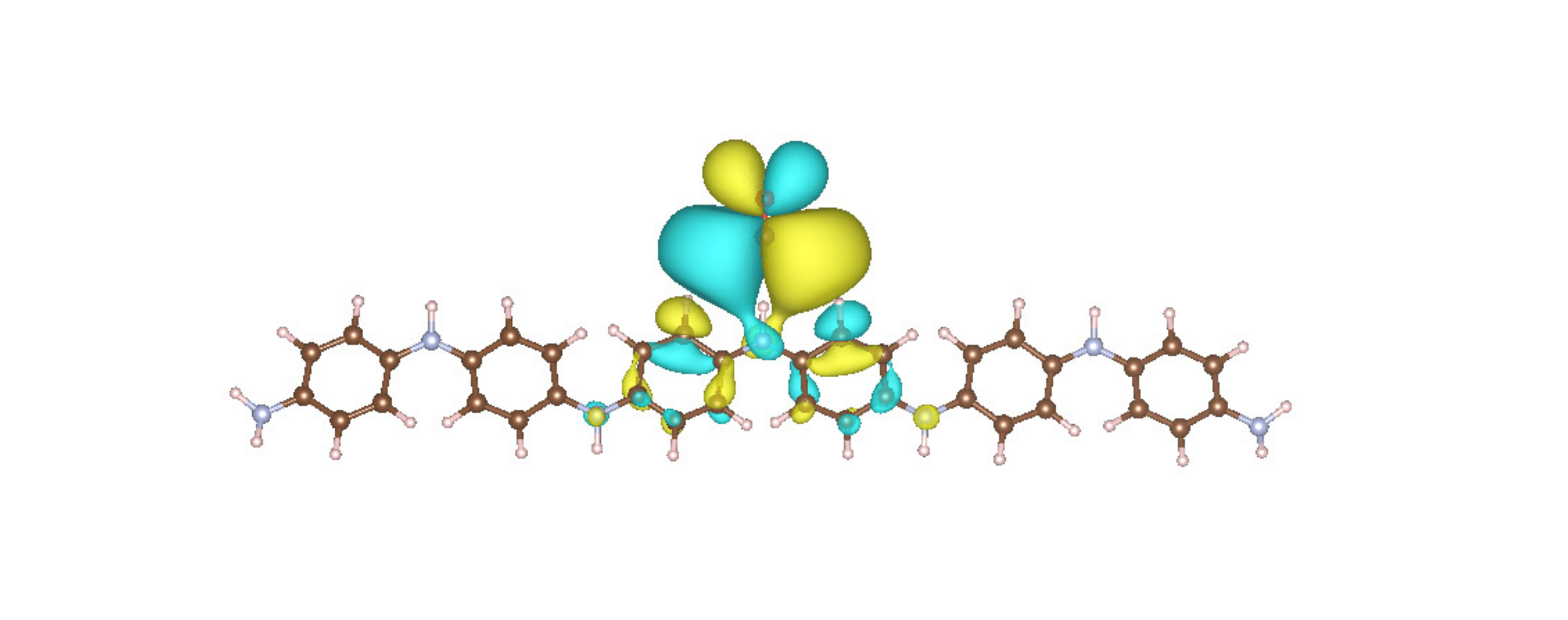} & \includegraphics*[scale=0.06]{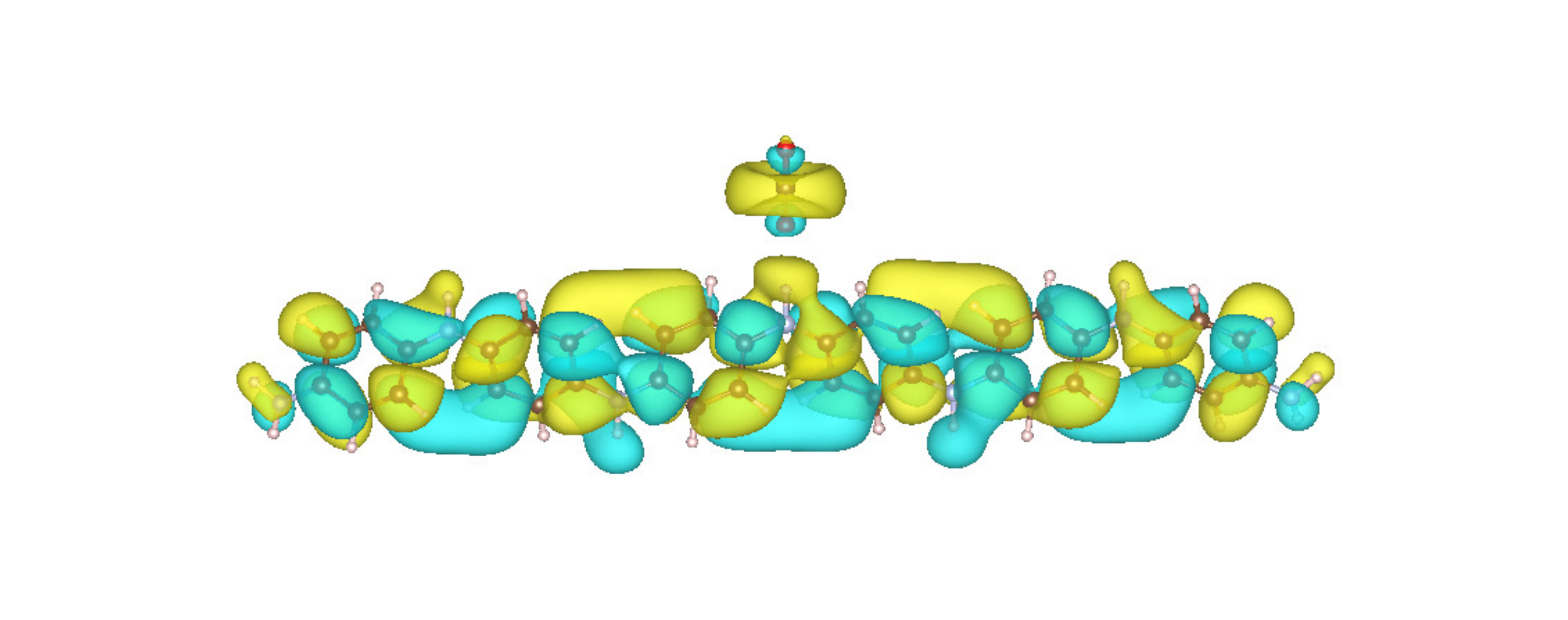} & \includegraphics*[scale=0.06]{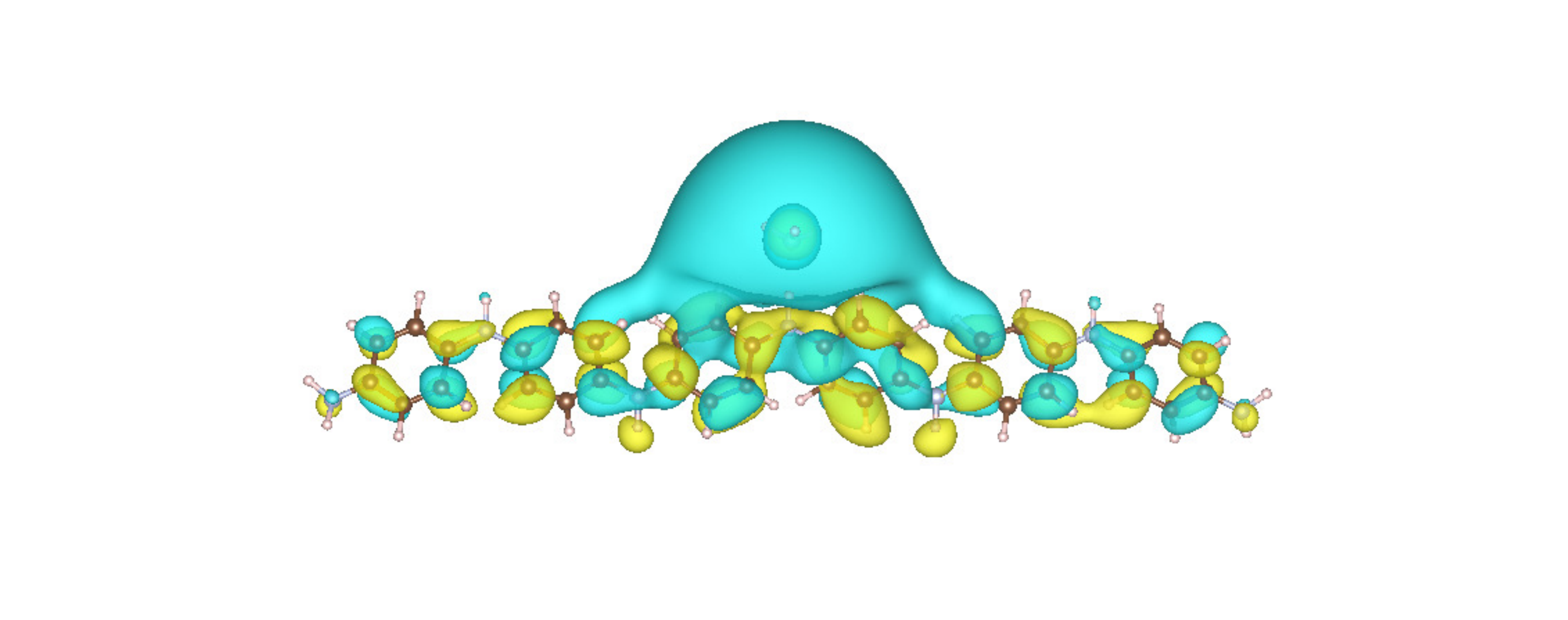} \\
HUMO       &  \includegraphics*[scale=0.06]{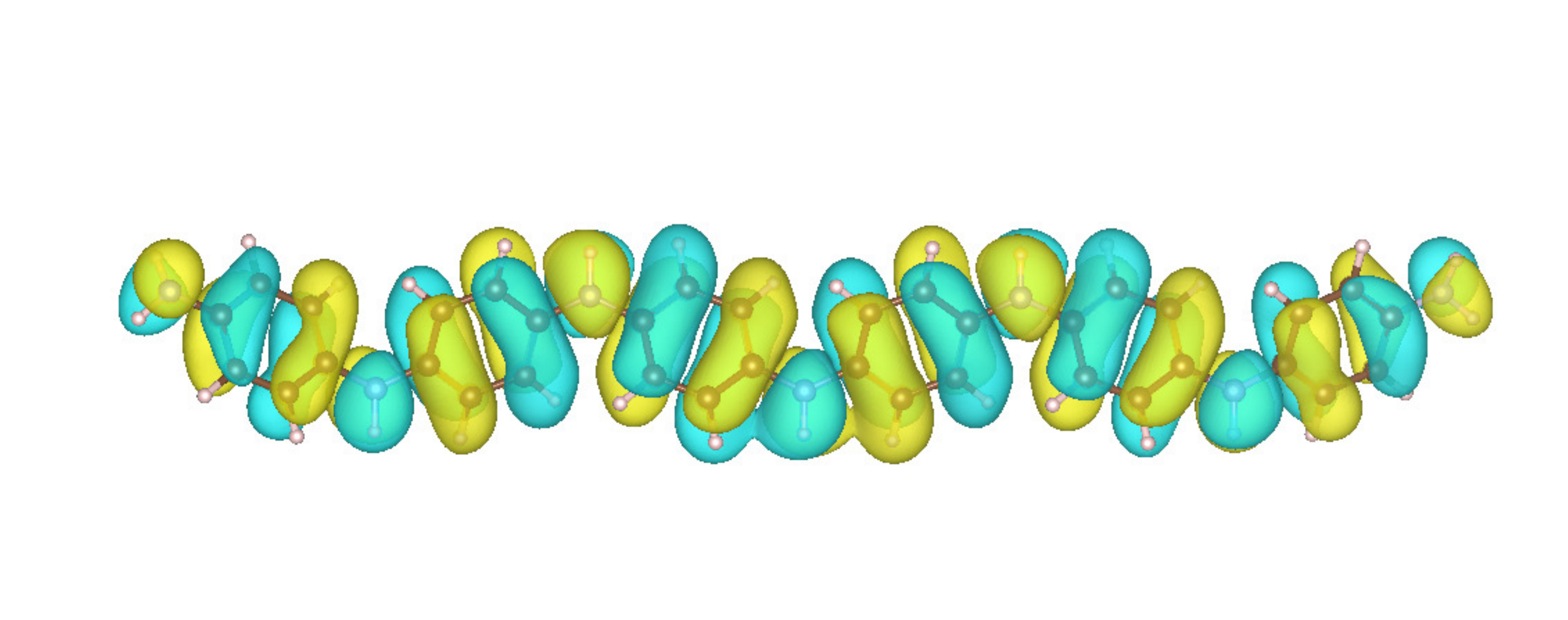} & \includegraphics*[scale=0.06]{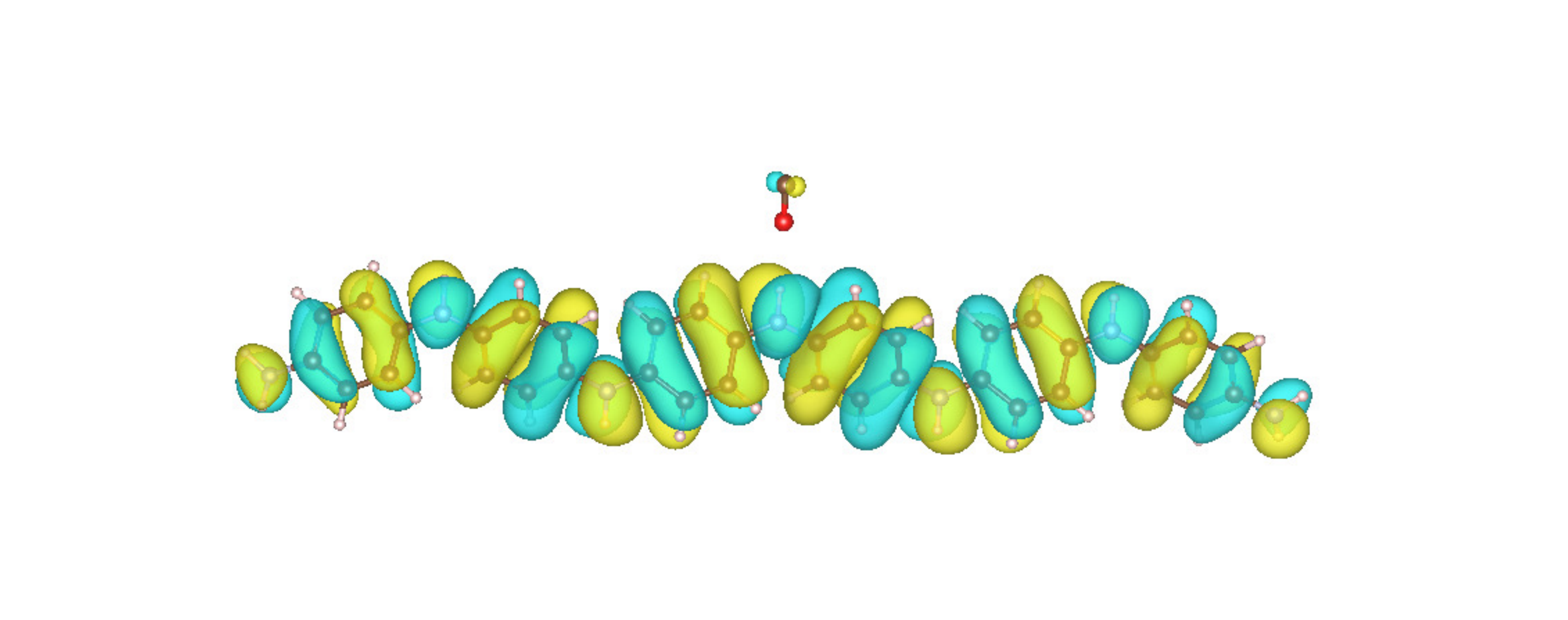}  & \includegraphics*[scale=0.06]{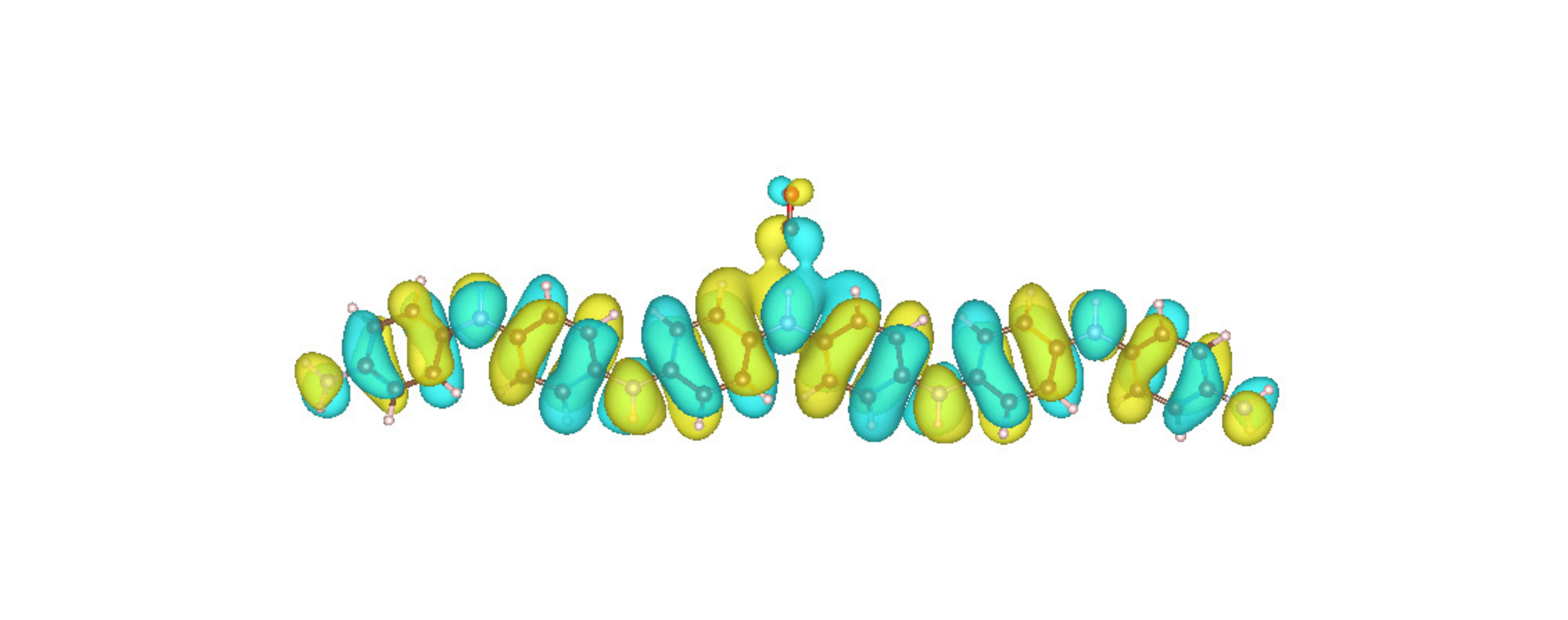} & \includegraphics*[scale=0.06]{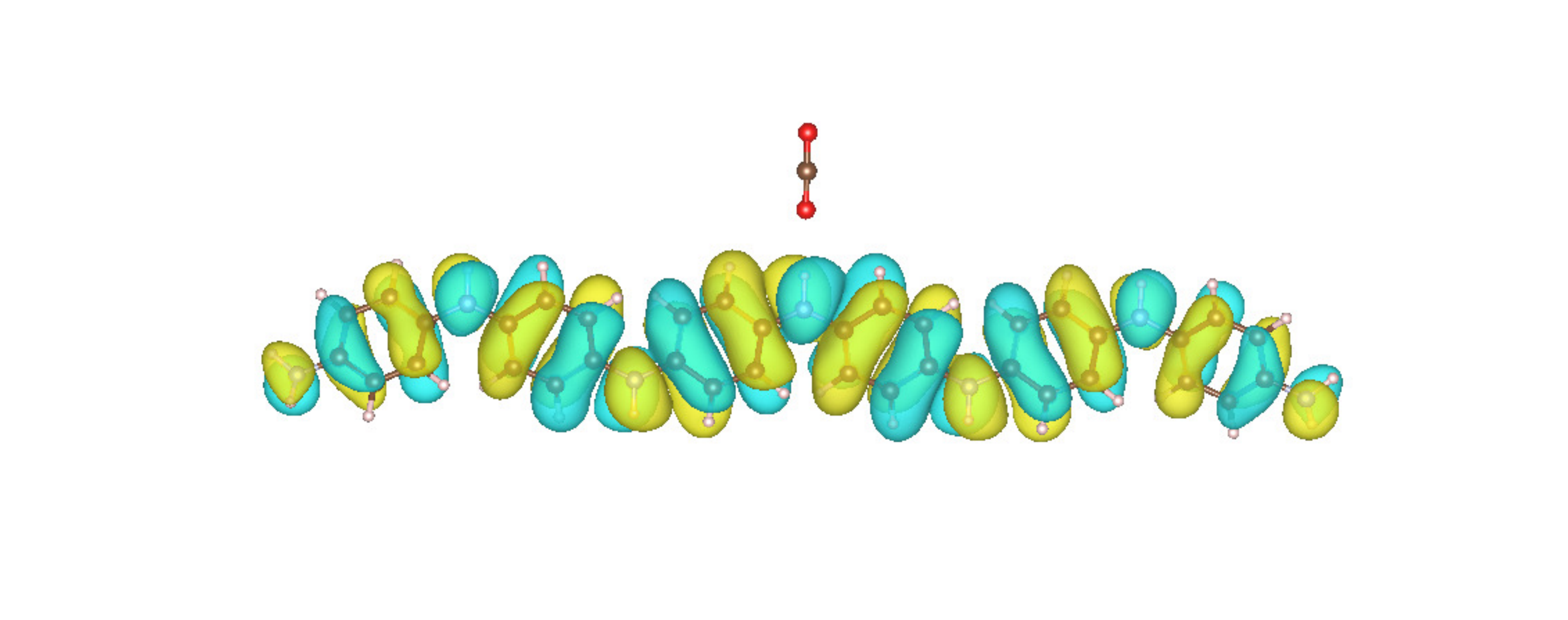} & \includegraphics*[scale=0.06]{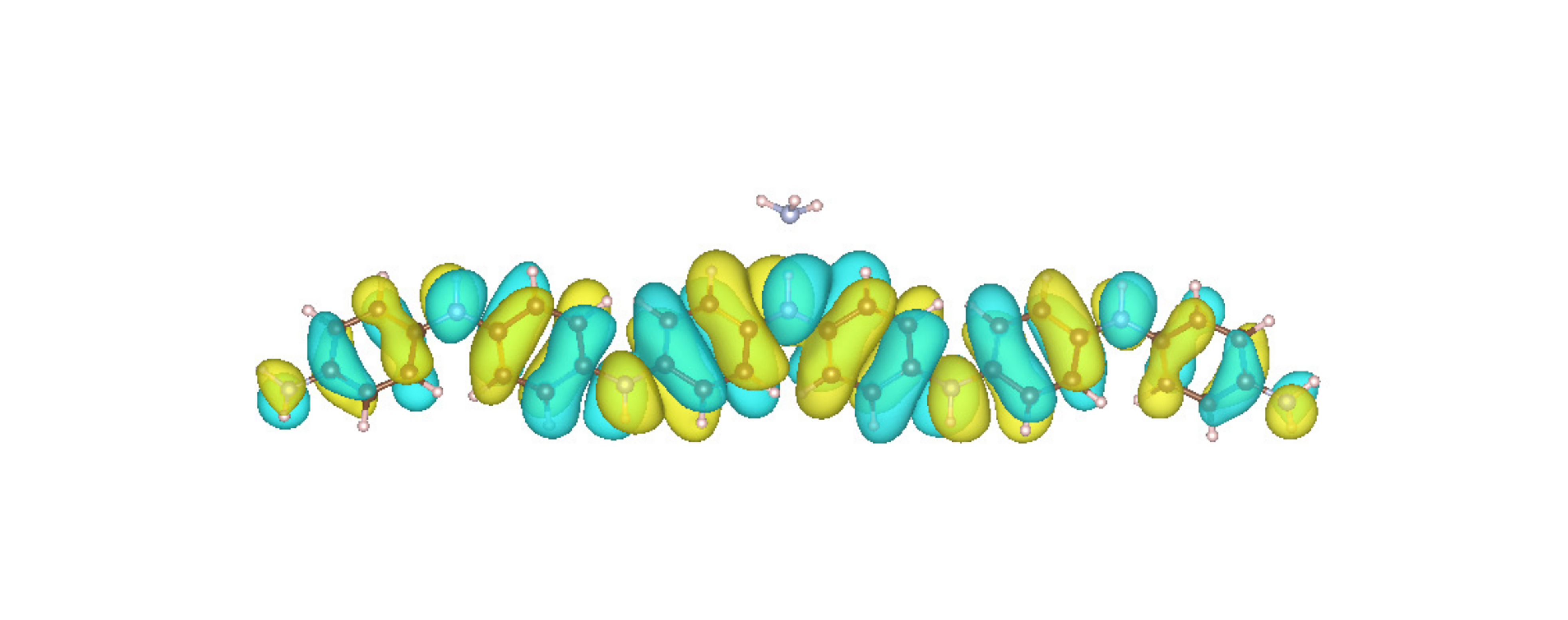} \\
\end{tabular}
\end{ruledtabular}
\end{table*}

Table \ref{geometry1} represents the global reactivity indices which are computed by DFT for
the highest stable configuration of gas molecules adsorbed onto nPANI ES molecule;
The electronic features of nPANI ES can also be discussed in terms of Ionization Potential (IP) and Electron Affinity (EA).
IP denotes the amount of energy required to remove the electron from nPANI ES. For n=2, high value of IP is observed for isolated nPANI ES, and  IP decrease after incorporation of gas molecules but high value of IP for n=4,6 relates to 4PANI ES-CO(1)and 6PANI ES-CO(2). High value of IP infers that the electrons in nPANI ES complexes are bound to the isolated nPANI ES . 
EA represents the alteration in the energy beacuse of the adding electron in nPANI ES. The high value of EA is 
one of the favorable conditions for chemical sensors.
Since the global hardness of a species is defined as its resistance against deformation in
the existance of an electric field, increasing the global hardness
increases the stability and reduces the reactivity of the species \cite{rocha2015ab}. 
The global electrophilicity includes data on the chemical
potential (which is related to electron transfer) and the hardness, (which is related to stability). 
Note that, for nPANI ES-CO(1) and nPANI ES-CO(2) the HL gap is relatively small, leading
to low hardness value for this complex comparing the softness , electrophilicity and electronic
chemical potential values which are high. It can be concluded that the enhanced hardness of complexes indicates that 
they are more stable in good according to the high HL gap recorded for these complexes.

\begin{table}[!ht]
\caption{\label{geometry1}
The global reactivity indices were calculated via DFT for
the most stable configuration of NH$_{3}$, CO(1), CO(2) and CO$_{2}$ adsorbed onto nPANI ES. 
IP: Ionizaton potential (-$E_{HOMO}$), EA: electron affinity (-$E_{LUMO}$), $\eta$ : hardness, $\mu$:  chemical potential,
S: softness.
}

\begin{ruledtabular}
\begin{tabular}{cccccc}
Species & IP(eV) & EA(eV) & $\eta$(eV) & $\mu$(eV) & S(eV) \\
\hline
2PANI ES           & 3.69 & 0.97 & 1.36  & -2.33 & 0.36 \\
2PANI ES-CO(2)     & 3.63 & 2.50 & 0.56 & -3.06 & 0.88 \\
2PANI ES-CO(1)     & 3.60 & 2.59 & 0.50 & -3.09 & 0.99  \\
2PANI ES-CO$_{2}$  & 3.56 & 1.15 & 1.20 & -2.35 & 0.41 \\
2PANI ES-NH$_{3}$  & 3.56 & 0.95 & 1.30 & -2.25 & 0.38 \\
4PANI ES           & 3.39 & 1.07 & 1.15 & -2.23 & 0.43 \\
4PANI ES-CO(2)     & 3.41 & 2.47 & 0.46 & -2.94 & 1.07 \\
4PANI ES-CO(1)     & 3.42 & 2.39   & 0.51 & -2.91 & 0.96  \\
4PANI ES-CO$_{2}$  & 3.37 & 1.06 & 1.15 & -2.22 & 0.43 \\
4PANI ES-NH$_{3}$  & 3.23 & 0.89 & 1.17 & -2.06 & 0.43 \\
6PANI ES           & 3.35 & 0.98 & 1.18  & -2.17 & 0.42 \\
6PANI ES-CO(2)     & 3.37 & 2.41 & 0.47 & -2.89 & 1.04  \\
6PANI ES-CO(1)     & 3.36 & 2.49 & 0.43 & -2.92 & 1.14 \\
6PANI ES-CO$_{2}$  & 3.33 & 1.79  & 1.13 & -2.20 & 0.45 \\
6PANI ES-NH$_{3}$  & 3.22 & 0.97 & 1.12 & -2.09 & 0.44 \\
\end{tabular}
\end{ruledtabular}
\end{table}

\subsection{Electronic and Vibrational Properties}

\begin{figure}[!ht]
\centering
\includegraphics*[scale=0.5]{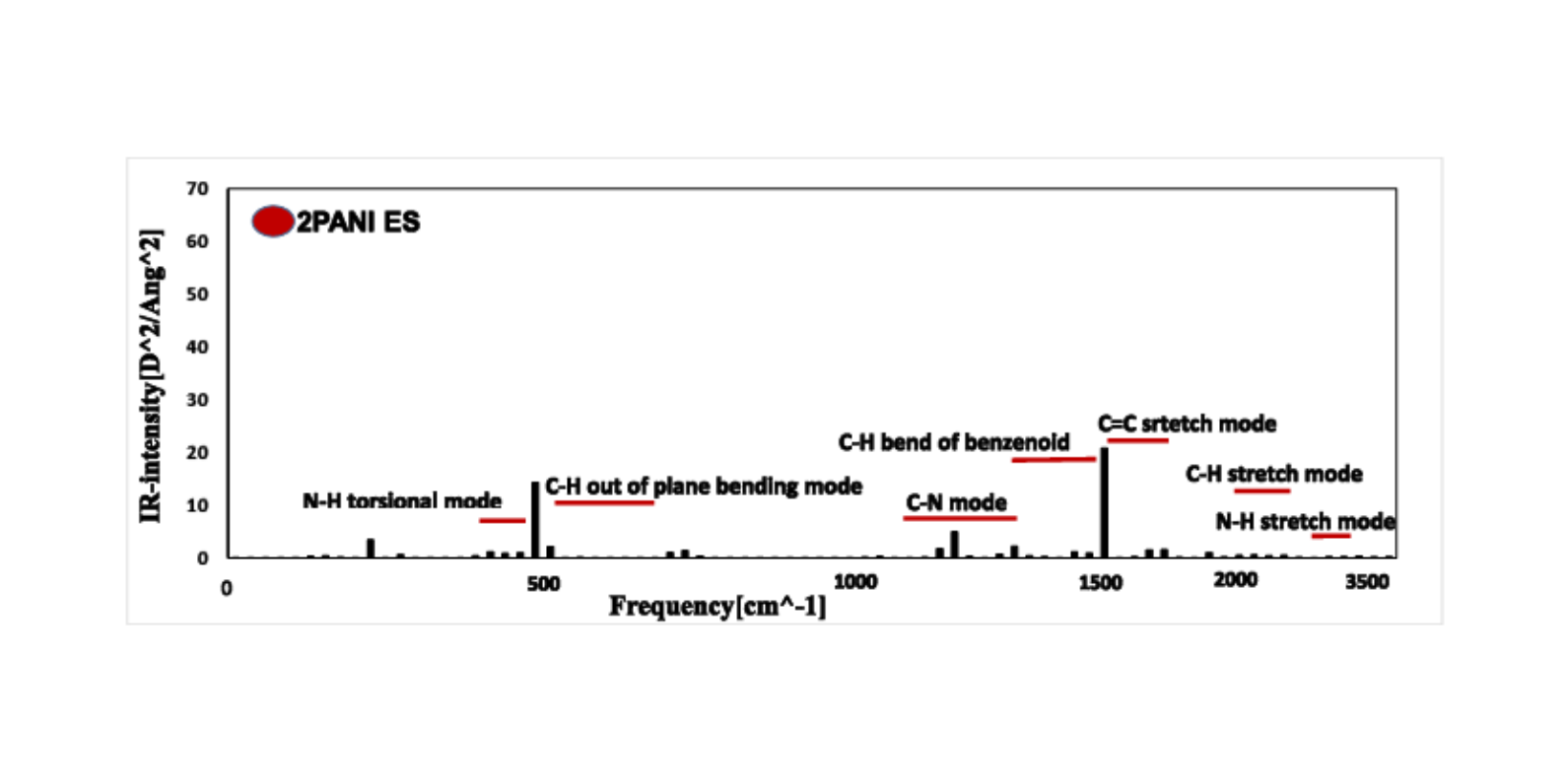}
\includegraphics*[scale=0.6]{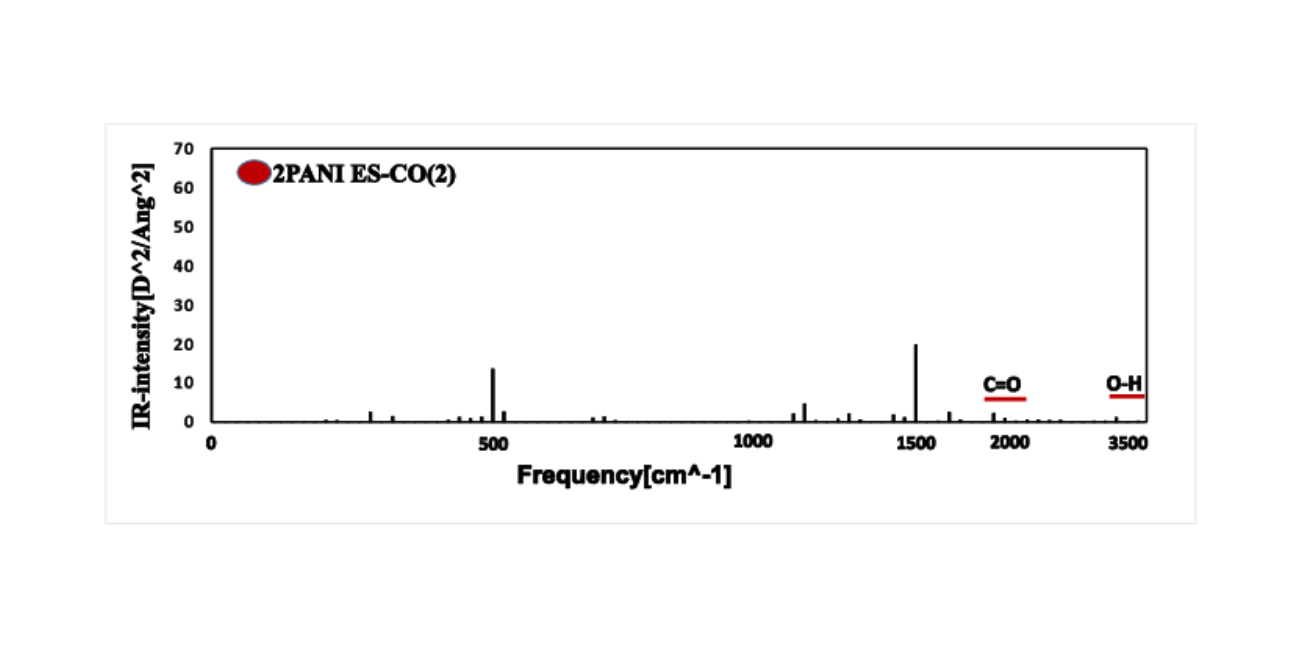}
\includegraphics*[scale=0.6]{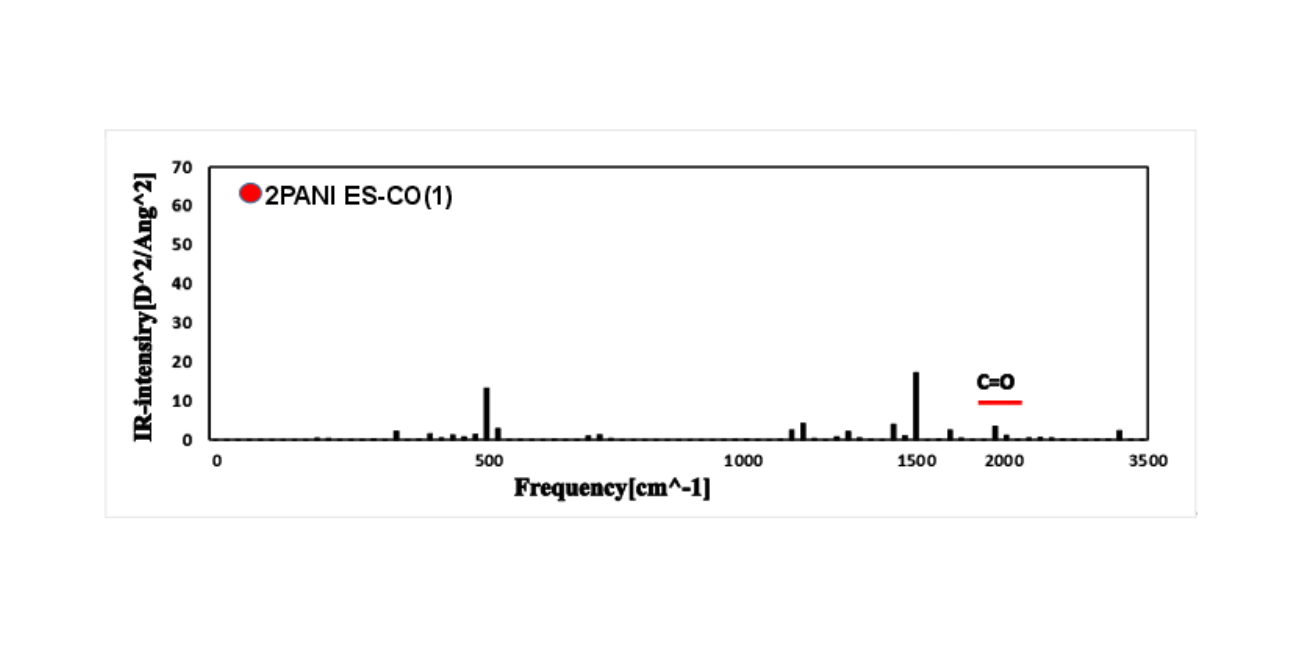}
\includegraphics*[scale=0.6]{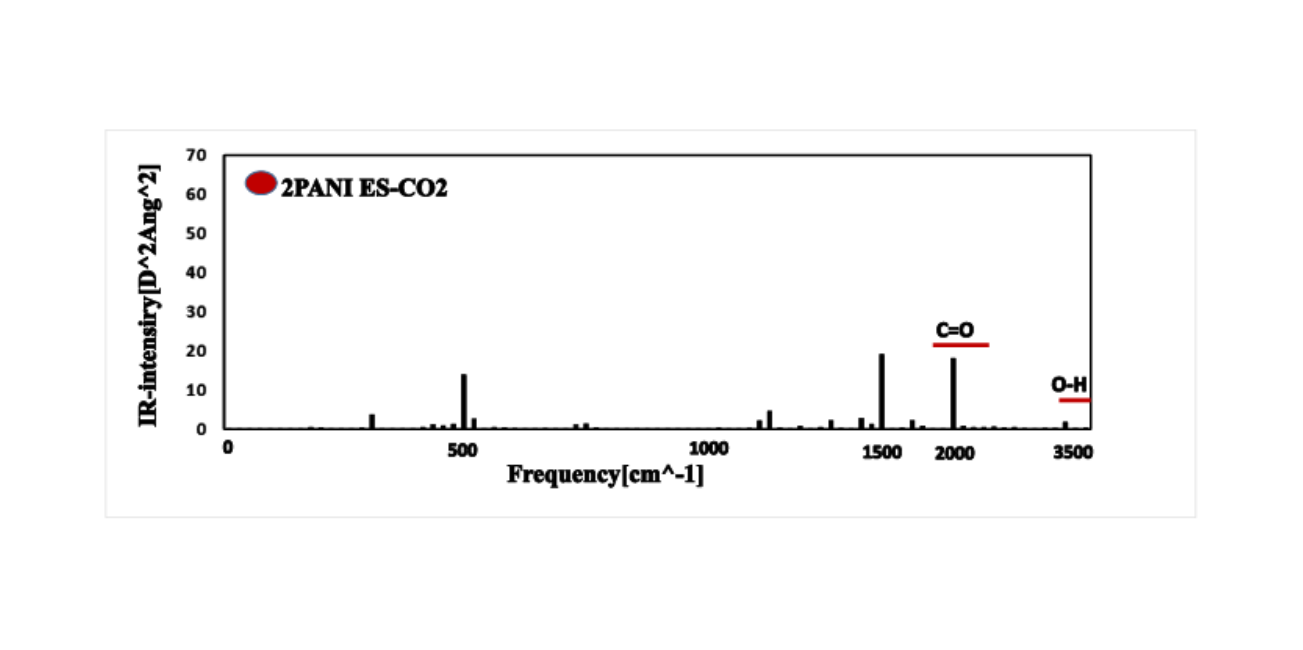}
\includegraphics*[scale=0.6]{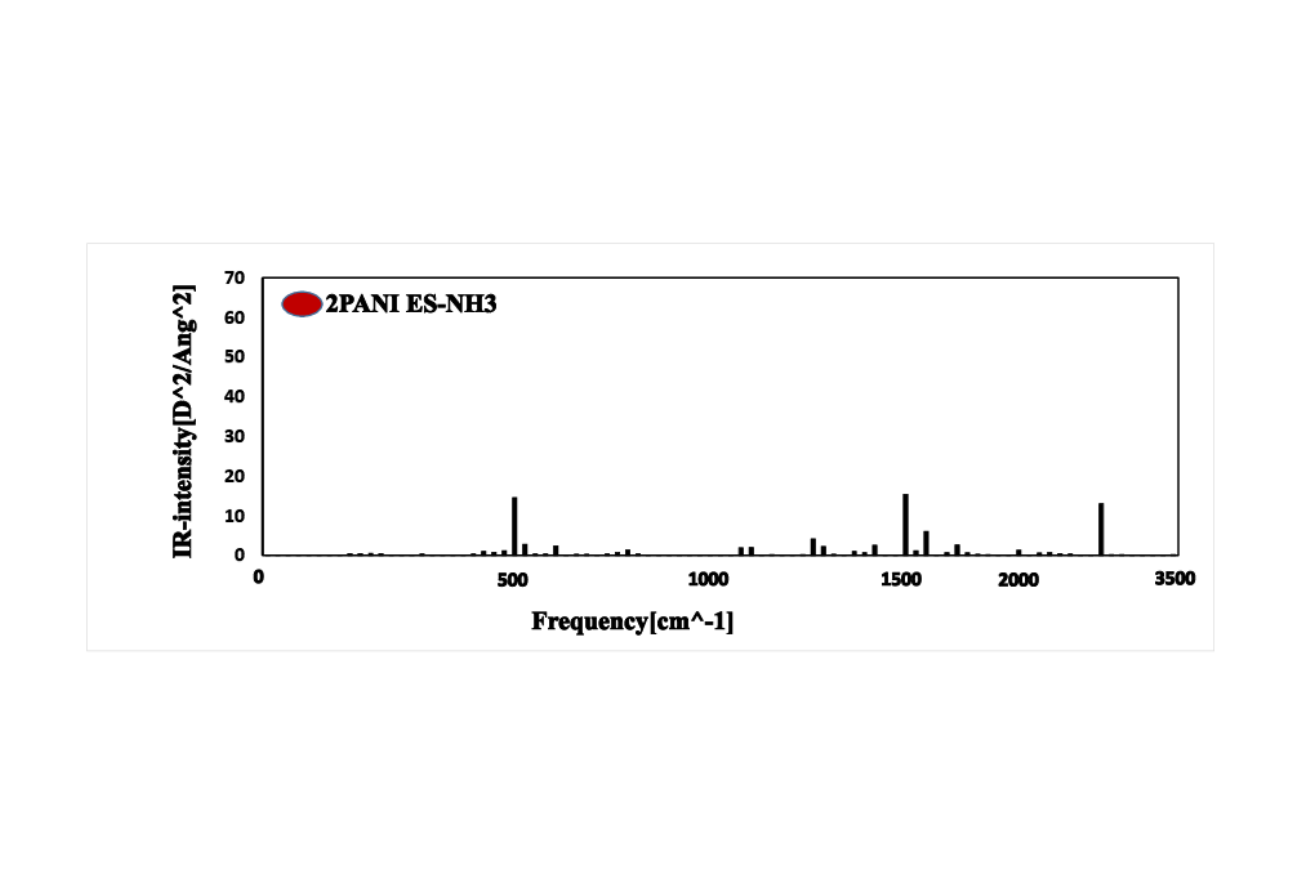}
\caption{\label{ir}
 calculated IR spectrum of 2PANI ES with gas molecules that adsorbed on it.
}
\end{figure}

\begin{figure}[!ht]
\centering
\includegraphics*[scale=0.6]{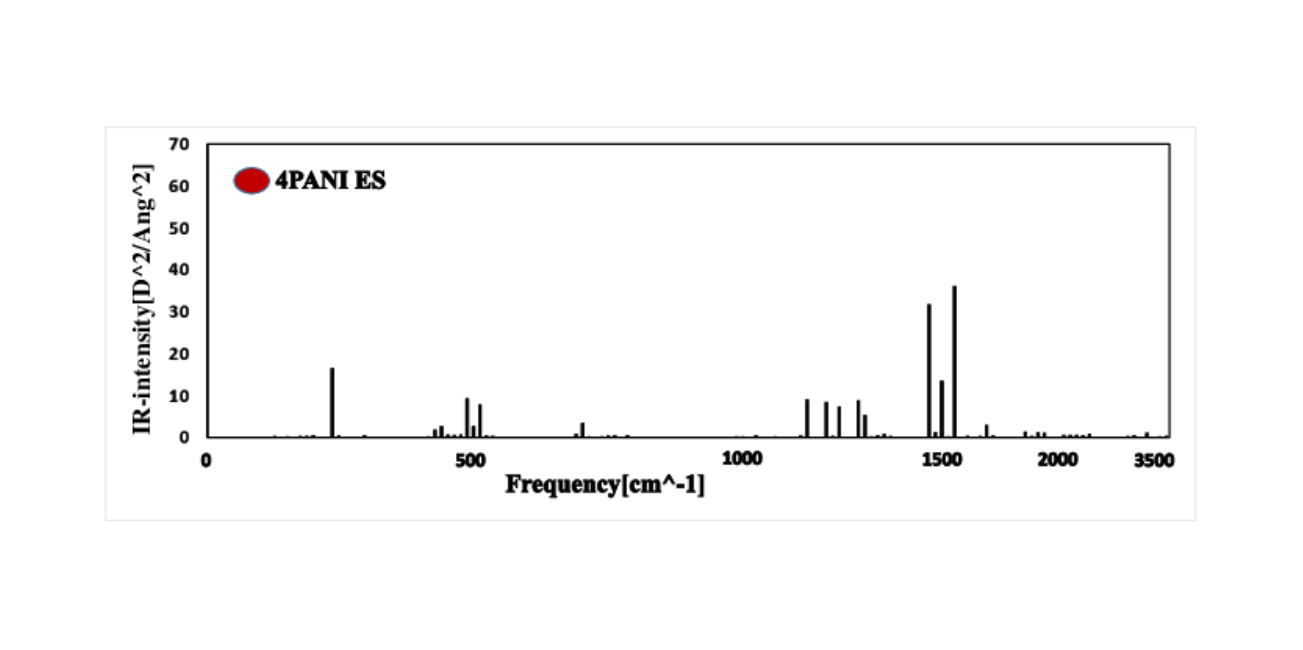}
\includegraphics*[scale=0.6]{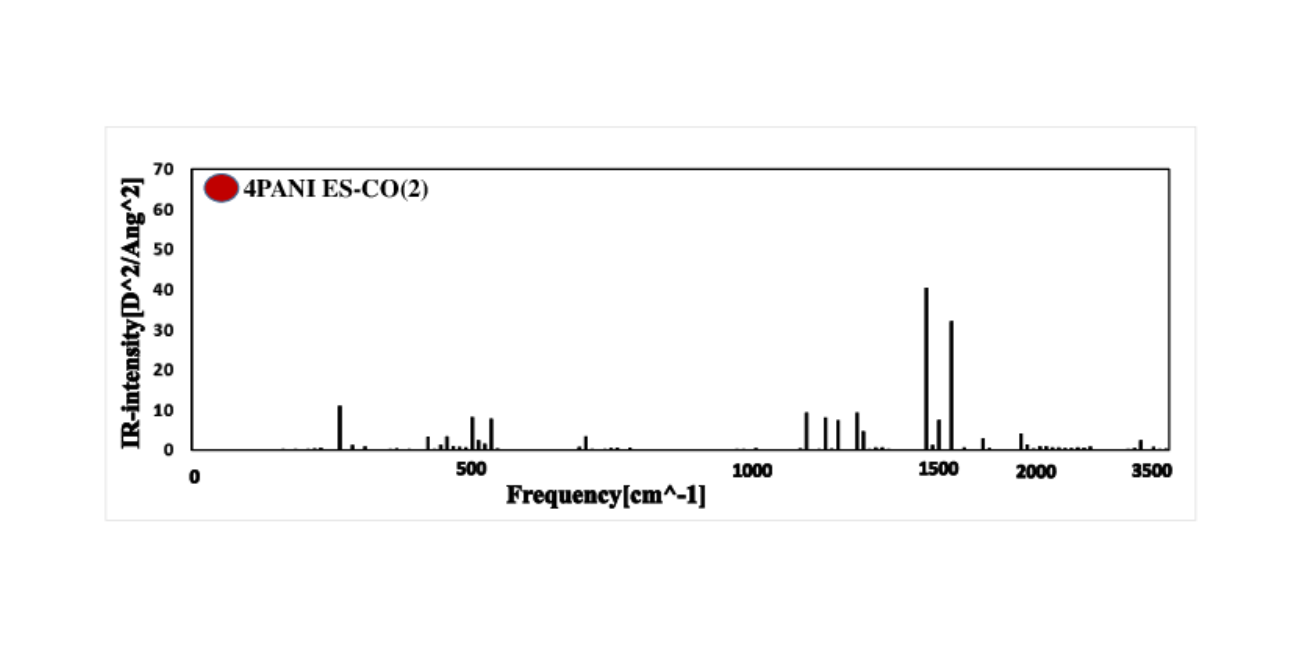}
\includegraphics*[scale=0.6]{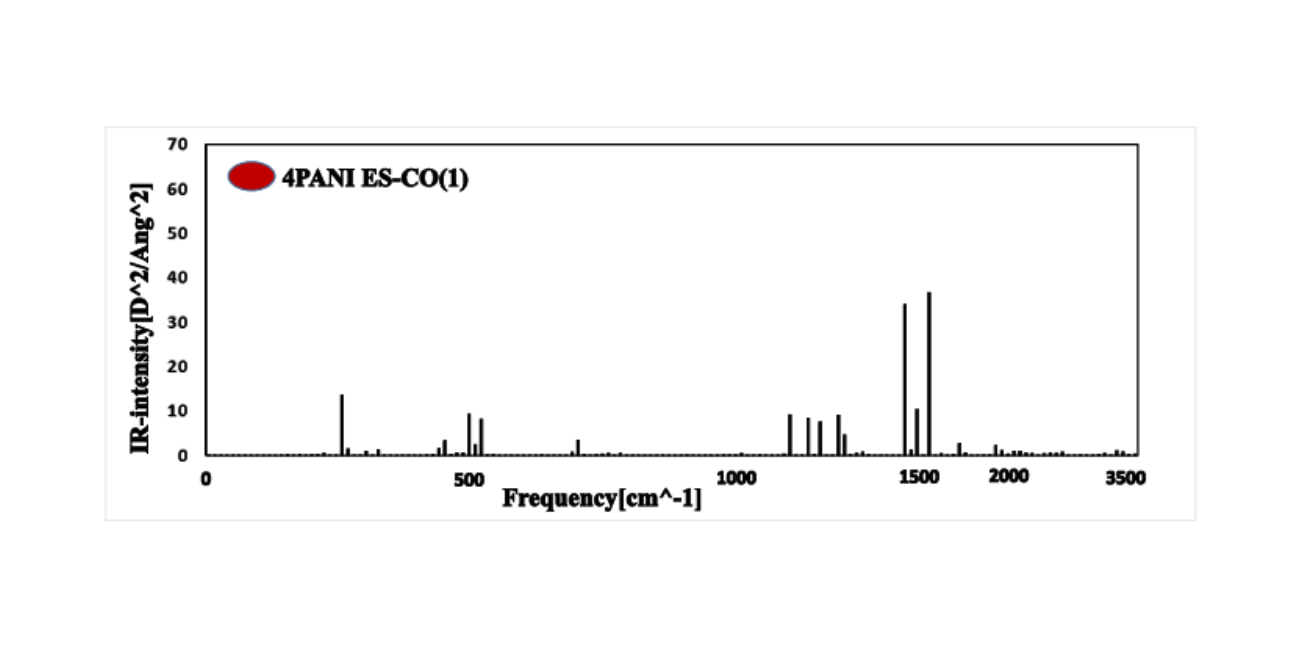}
\includegraphics*[scale=0.6]{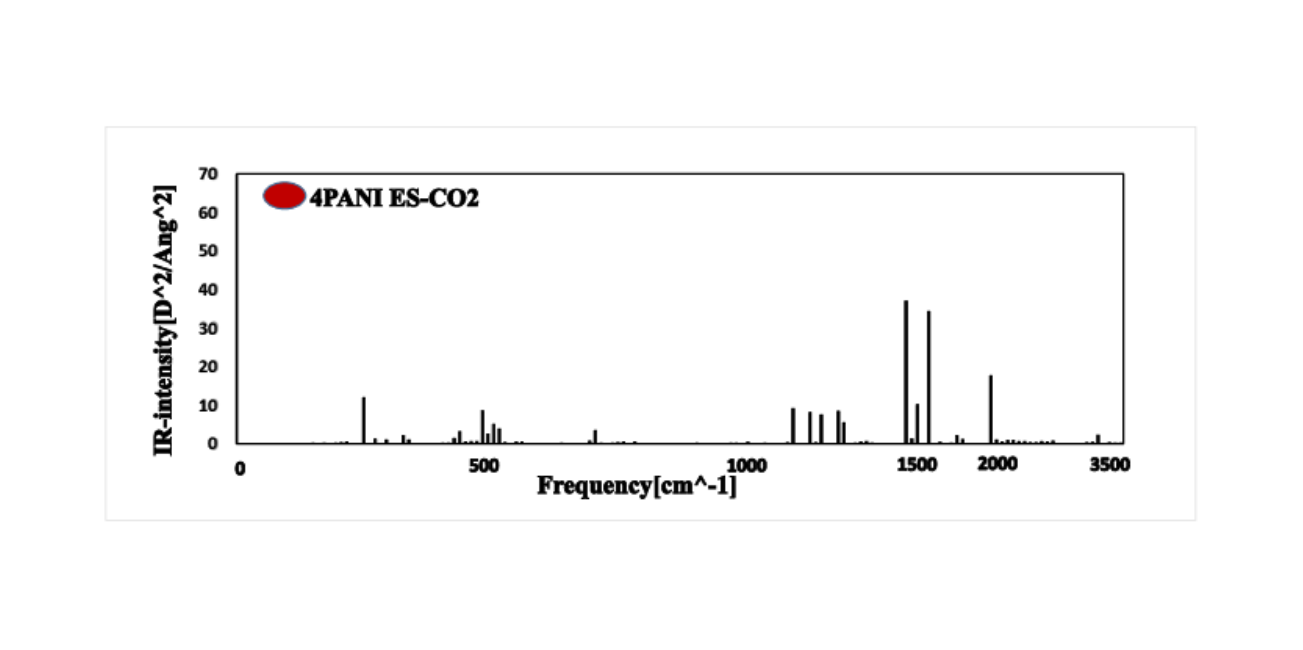}
\includegraphics*[scale=0.6]{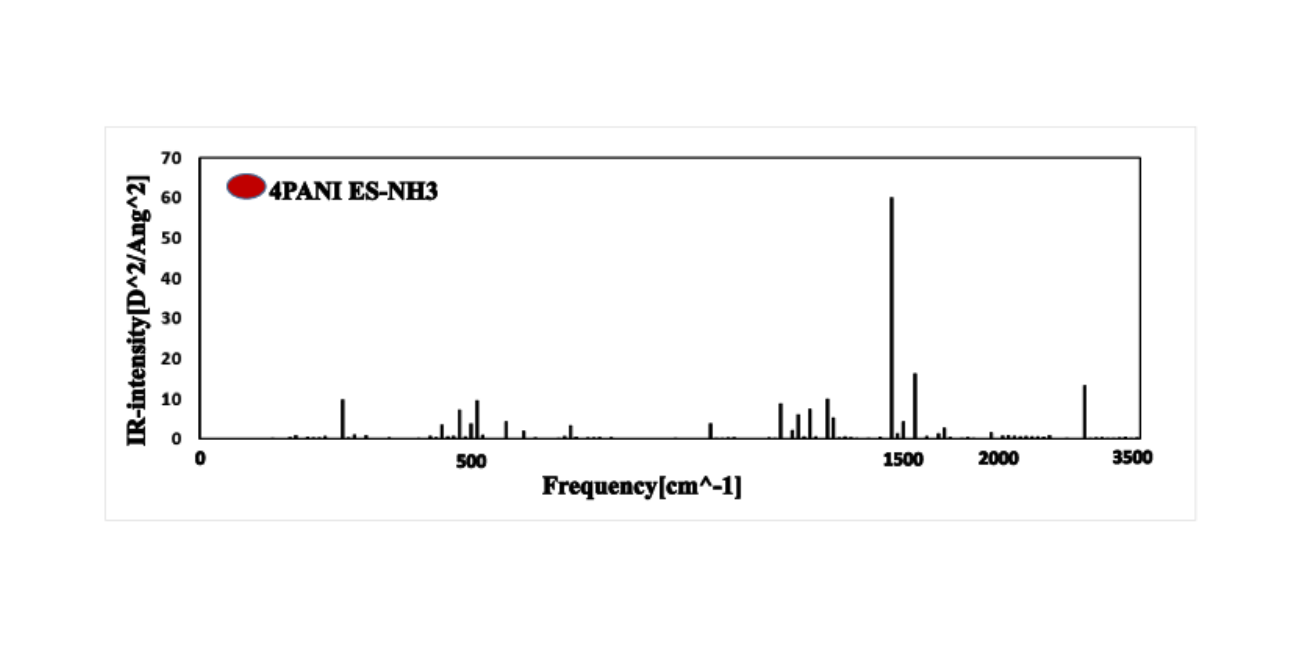}
\caption{\label{ir1}
 calculated IR spectrum of 4PANI ES with gas molecules that adsorbed on it.
}
\end{figure}

\begin{figure}[!ht]
\centering
\includegraphics*[scale=0.6]{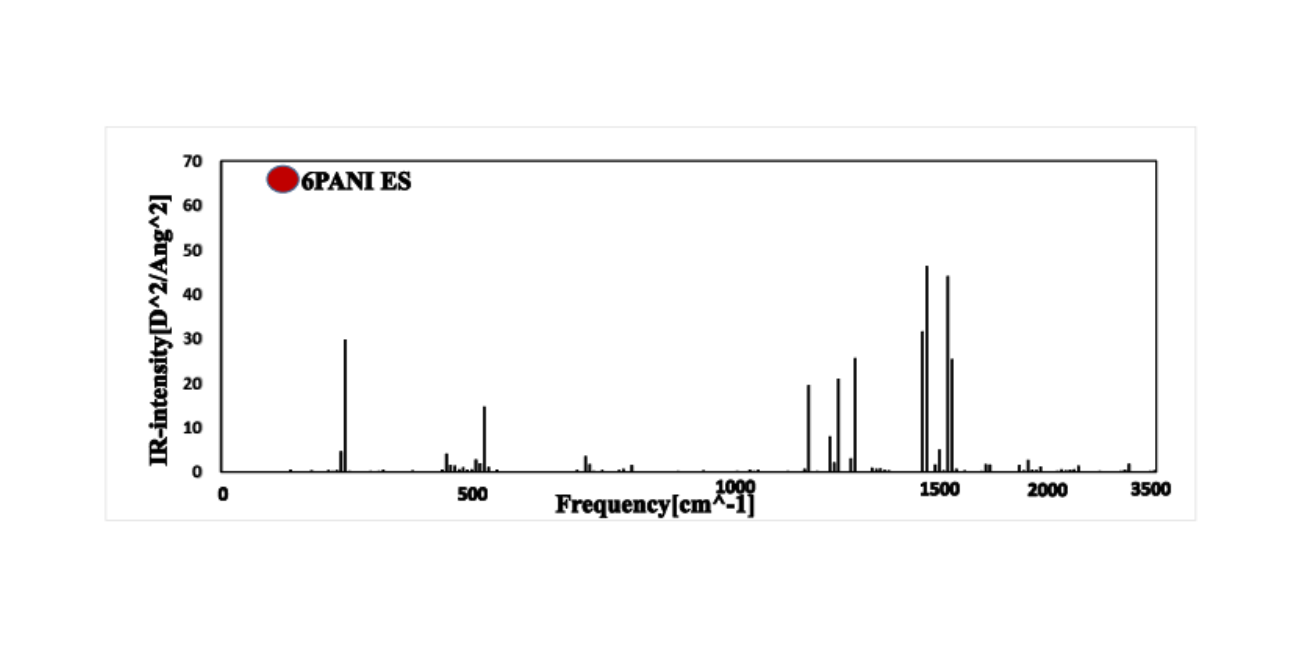}
\includegraphics*[scale=0.6]{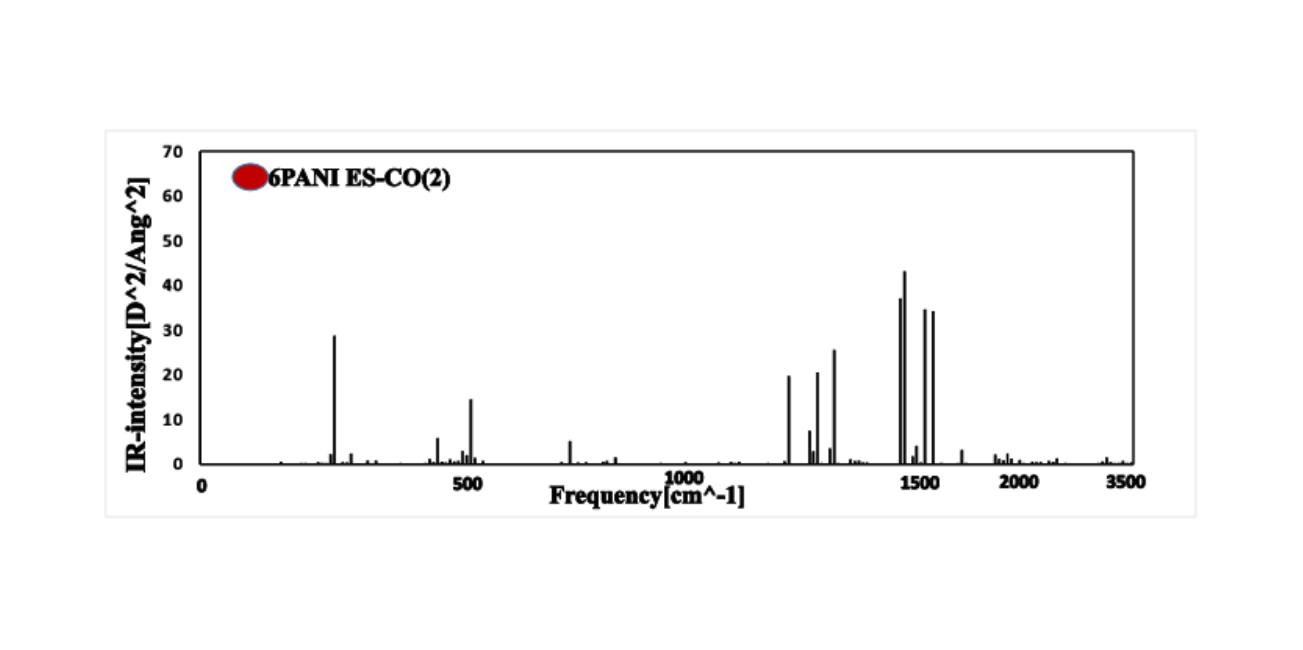}
\includegraphics*[scale=0.5]{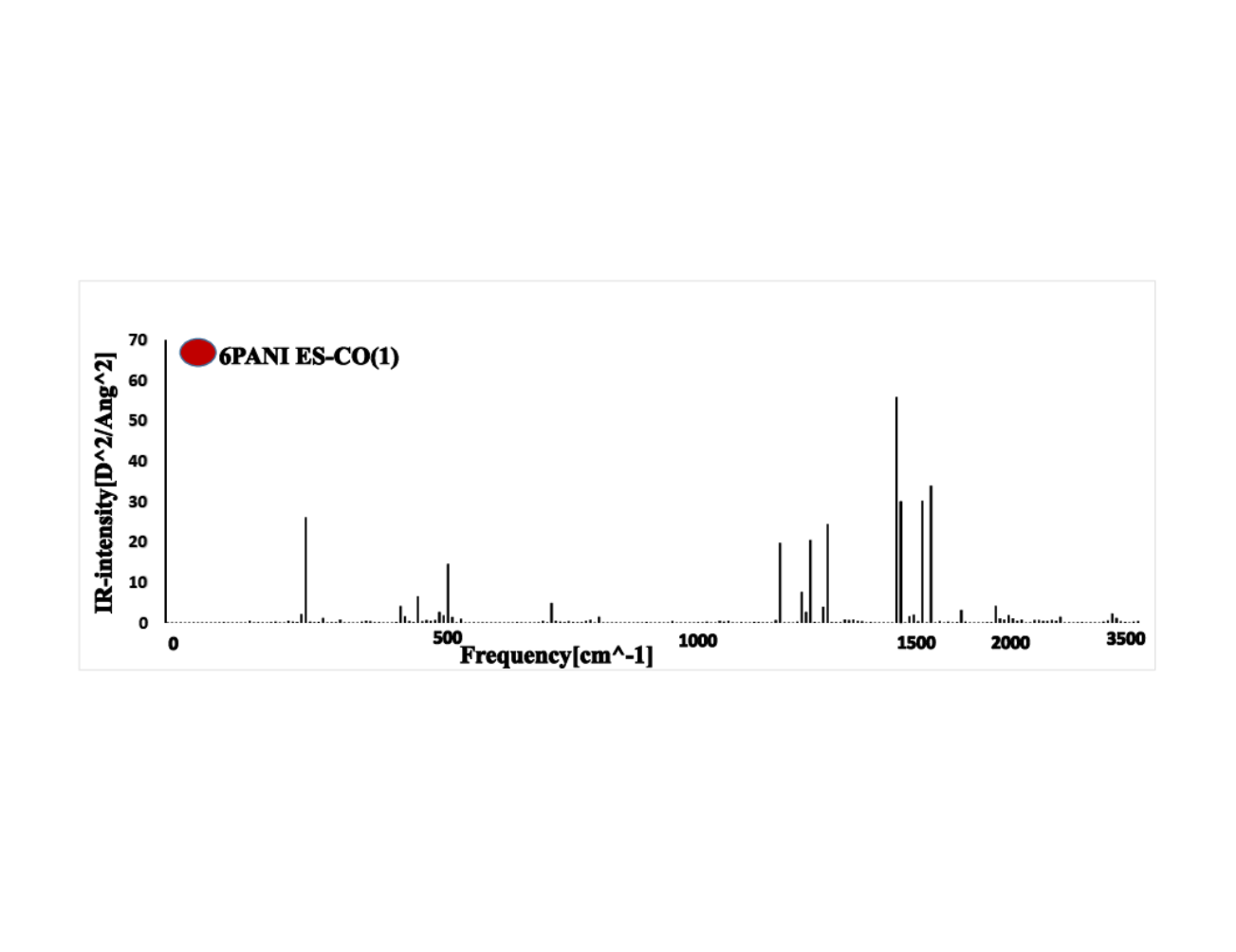}
\includegraphics*[scale=0.6]{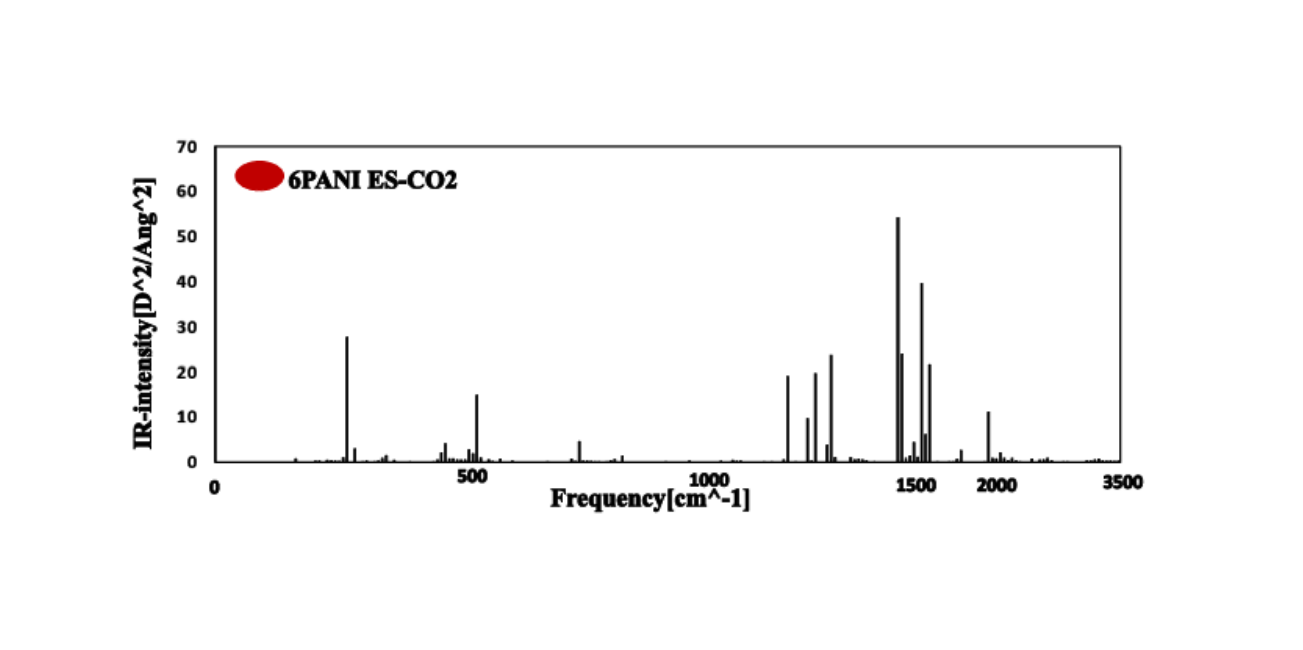}
\includegraphics*[scale=0.6]{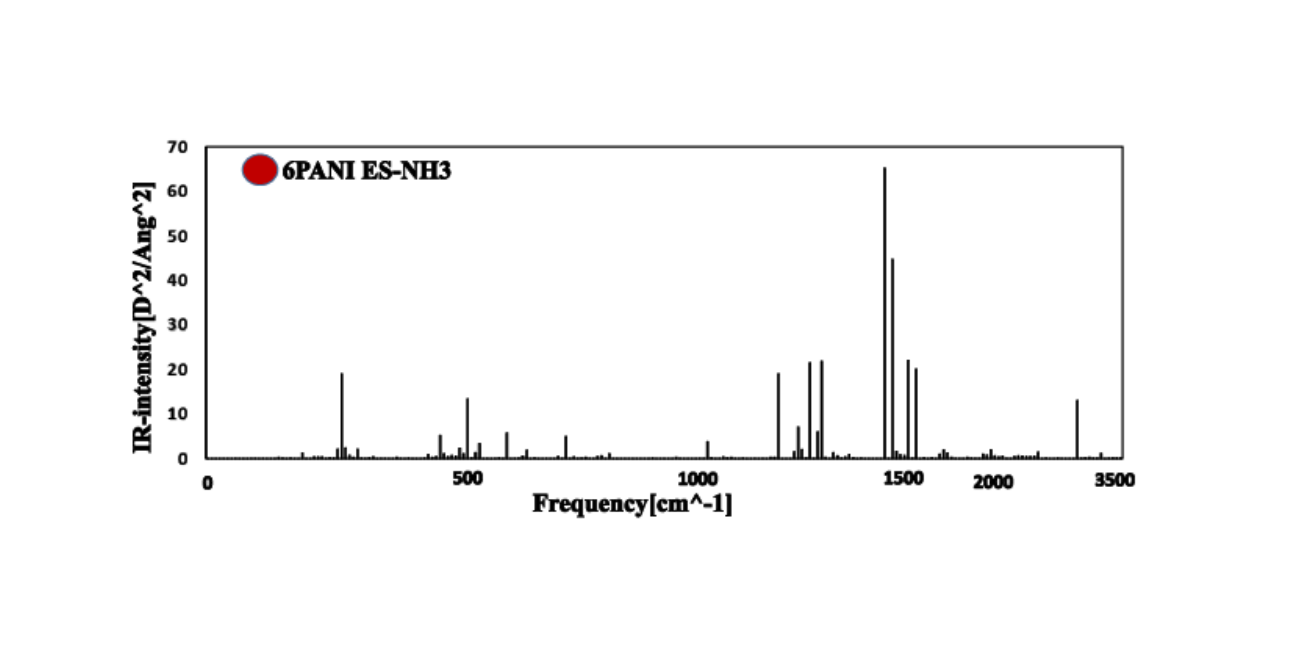}
\caption{\label{ir2}
 calculated IR spectrum of 6PANI ES with gas molecules that adsorbed on it.
}
\end{figure}

The electronic structures corresponding to the optimized nPANI ES and nPANI ES-X molecules are calculated 
by using both pseudo-potential and full potential methodes with a good agreement with each other.
The HL gap of 2PANI ES was obtained 2.72 eV within pseudo-potential method, which is 
significantly bigger than the experimental gap of about 3.48 eV \cite{al2016effect}.
This difference shows the major deficiency of LDA/GGA functional for predicting  the excited states features.
Thus,TD-DFT calculations will be utilized to attain some reliable outcomes for HL gap.
The vibrational spectra corresponding to the molecule is computed to study dynamical stability and IR spectrum for the systems.
The resulted IR spectrum of the nPANI ES and nPANI ES-X molecules are listed in Fig. \ref{ir}, Fig. \ref{ir1} and Fig. \ref{ir2}. 
Dynamical stability of the molecule proved by the absence of any imaginary style in the spectrum.
As an evident, 2PANI ES, 4PANI ES and 6PANI ES have 3 greater IR intensity that relate to the stronger bonds.
Vibrations around 598 $(cm^{-1})$, 1508 $(cm^{-1})$, 1264 $(cm^{-1})$, 296 $(cm^{-1})$, 1506 $(cm^{-1})$, 
1509 $(cm^{-1})$, 288 $(cm^{-1})$, 1312 $(cm^{-1})$ and 1482 $(cm^{-1})$ frequency show greater 
IR intensity respectively for 2PANI ES, 4PANI ES and 6PANI ES.
We can conclude when rings of PANI ES increased from n=2 to n= 6, the value and position of IR
intensity, and vibration frequencies are changed.

In addition, when gas molecules get absorbed on nPANI ES, the value of IR intensity and vibrational frequencies are changed.
Interaction 2PANI ES-X with gas molecules where X=CO(1), CO(2), CO$_{2}$ causes new bonds C=O and O=H
around 2120 to 2355 $(cm^{-1})$ and around 3539 to 3564 $(cm^{-1})$, respectively. There is also change in the vibrational frequencies position with
higher IR intensity in 2PANI ES-X complexes compare to isolated 2PANI ES.
The highest amount of C=O bonds IR intensity occurs in nPANI ES-CO$_{2}$, while interacting NH$_{3}$ with 2PANI ES results in the biggest change in the vibrational frequencies. 
Similarly, the interaction of 4PANI ES-X and 6PANI ES-X (X=CO(1), CO(2), CO$_{2}$) with gas molecules causes 
new bonds C=O and O=H around 2120 to 2355 $(cm^{-1})$ and around 3539 to 3564 $(cm^{-1})$ respectively.
We can observe for 2PANI ES-NH$_{3}$, 4PANI ES-NH$_{3}$ and 6PANI ES-NH$_{3}$ the amount of a higher and increased IR intensity.
Analyzing the IR spectrum corresponding to the systems with adsorption of gas molecules on an isolated nPANI ES  clearly suggests that nPANI ES can be used as sensor.

\subsection{Optical properties}

\begin{figure}[!ht]
\centering
\includegraphics [scale=0.22]{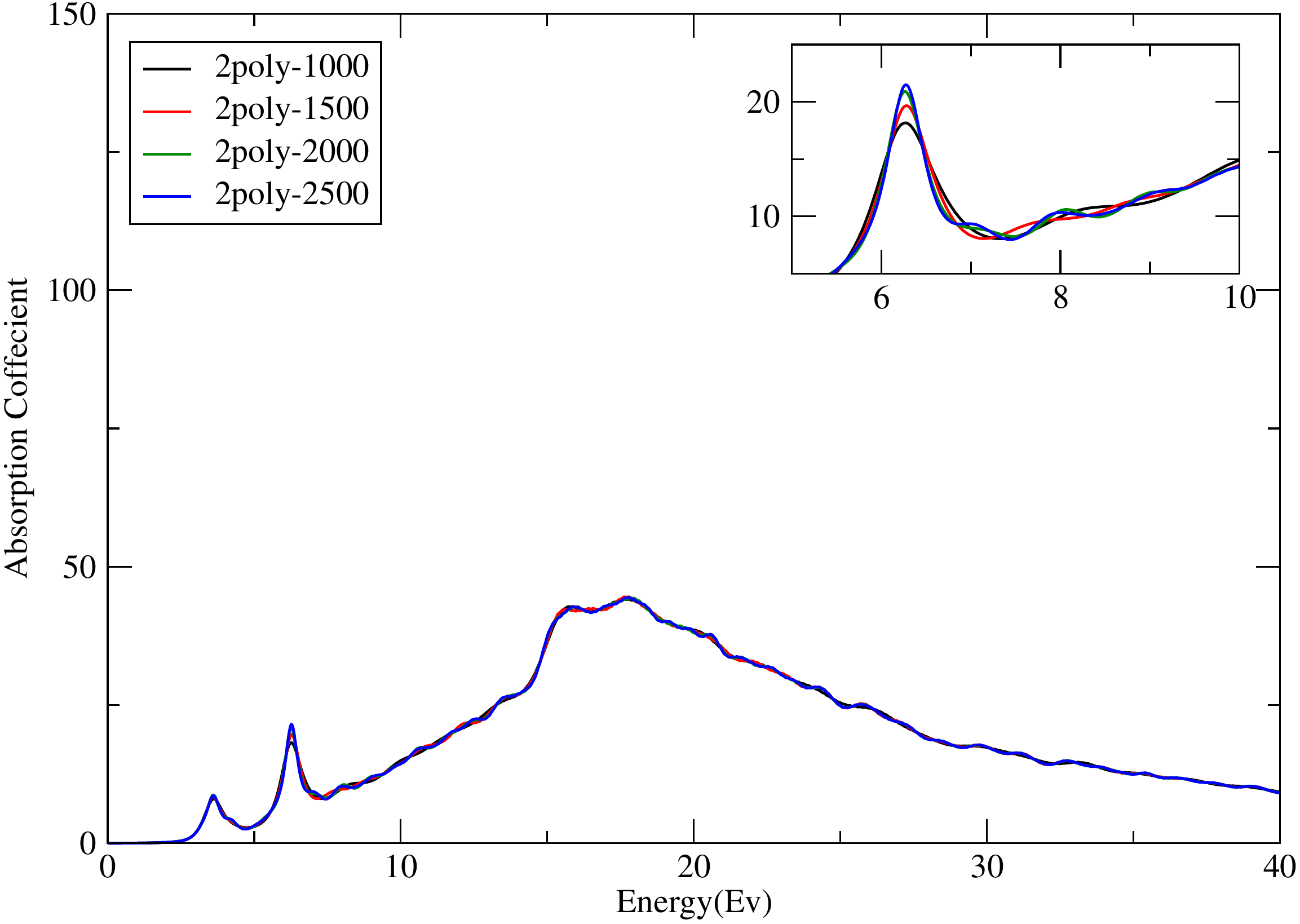}
\caption{\label{iter}
  Optimizing the number of iterations to determine 
  the optical spectrum of the 2PANI ES molecule.
}
\end{figure}

The Liovile-lanczos approach 
with TD-DFT is used for determining the optical spectrum. 
In this method, the optimization of iterations number should be performed.
The absorption spectrum corresponding to 2PANI ES have been calculated
using various iterations number (Fig.~\ref{iter}). 
It is observed that a flat optical spectrum  is resulted with 1000 iteration, 
while sharp peaks is observed by enhancing iterations. 
The obtained absorption spectra with 2000 and 2500 iterations are fully overlapped. 
Therefore, 2000 iterations is the optimal factor for our following computations.

\begin{figure}[!ht]
\centering
\includegraphics*[scale=0.3]{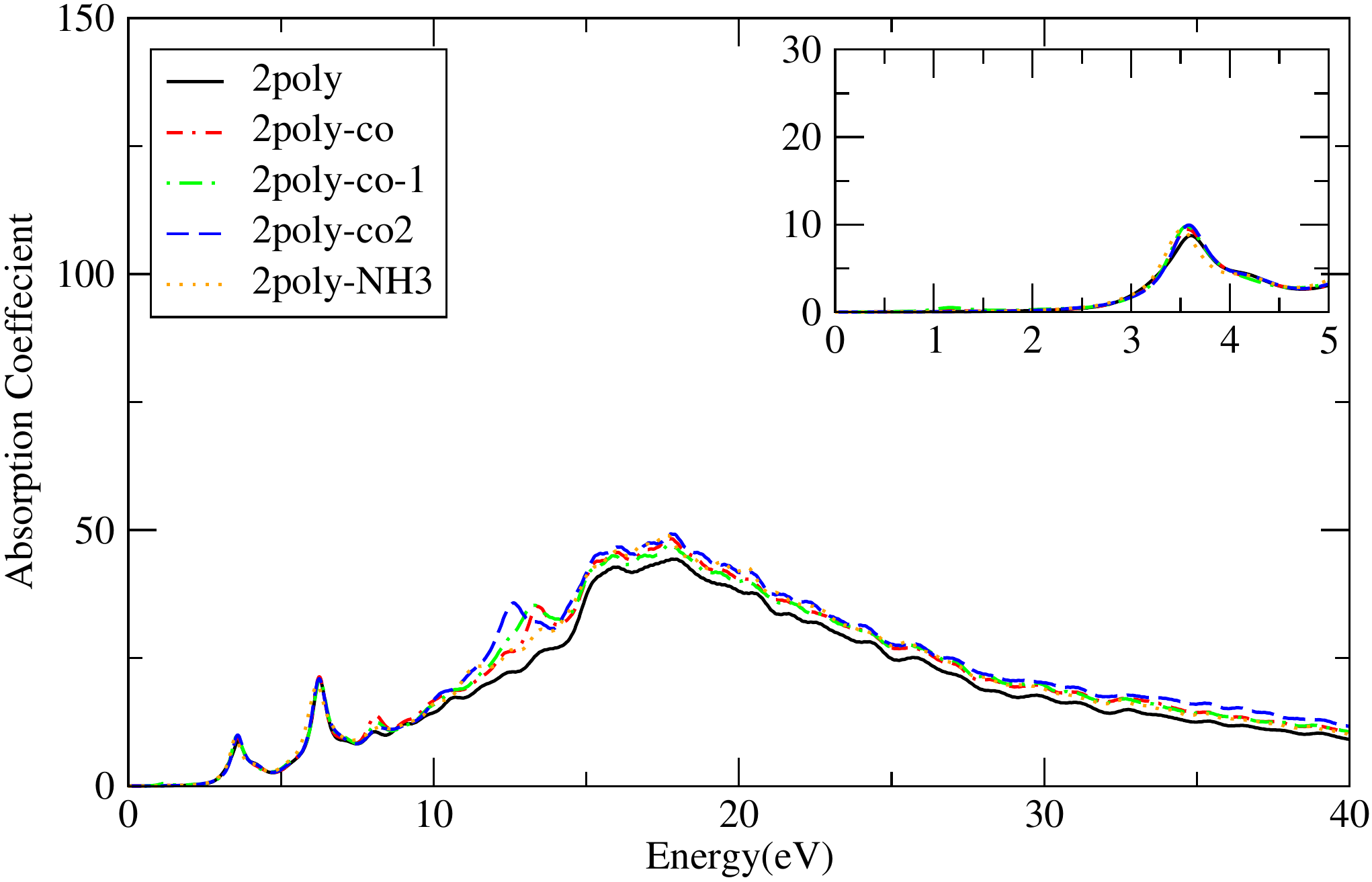}
\includegraphics*[scale=0.3] {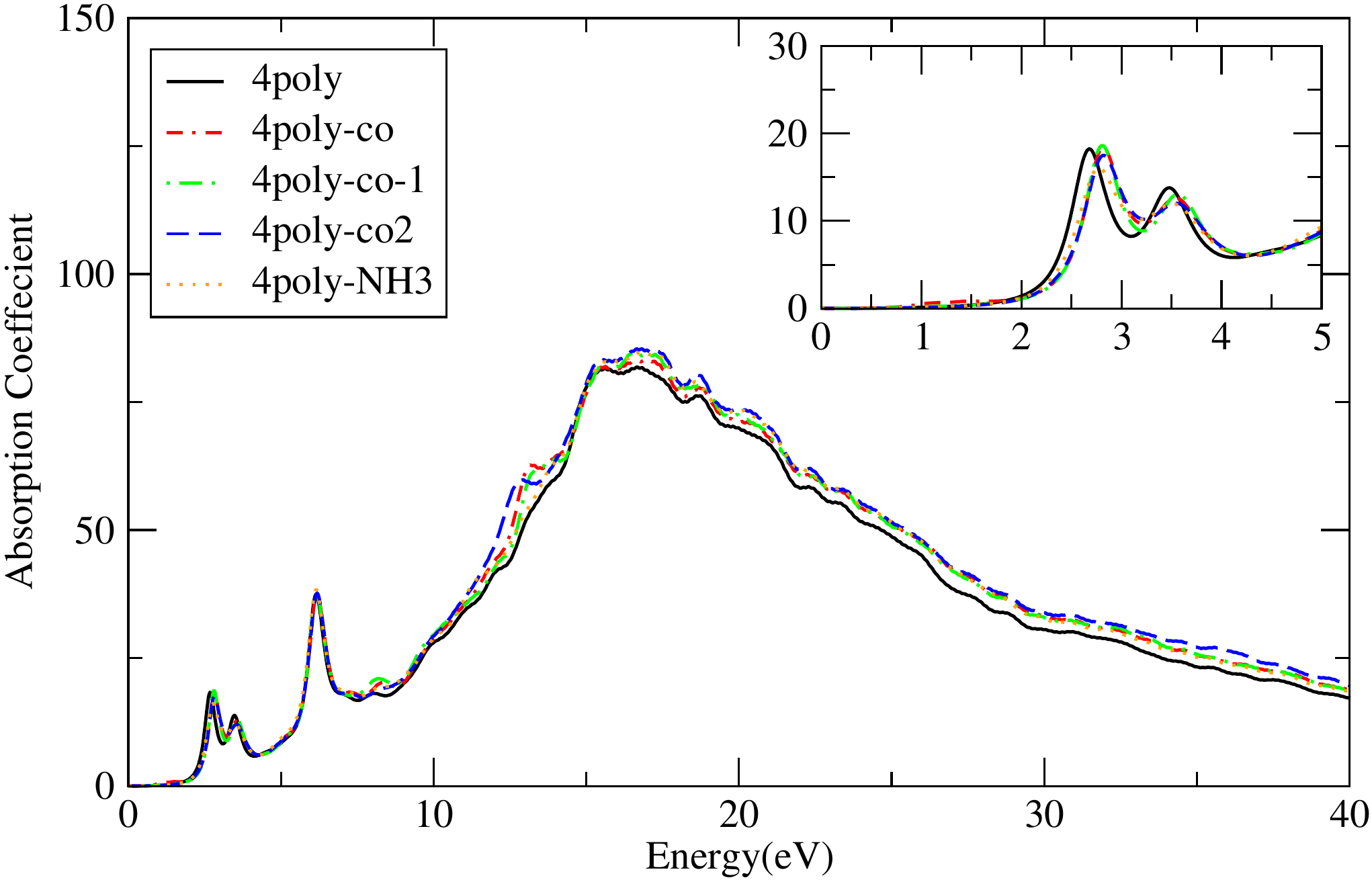}
\includegraphics*[scale=0.3]{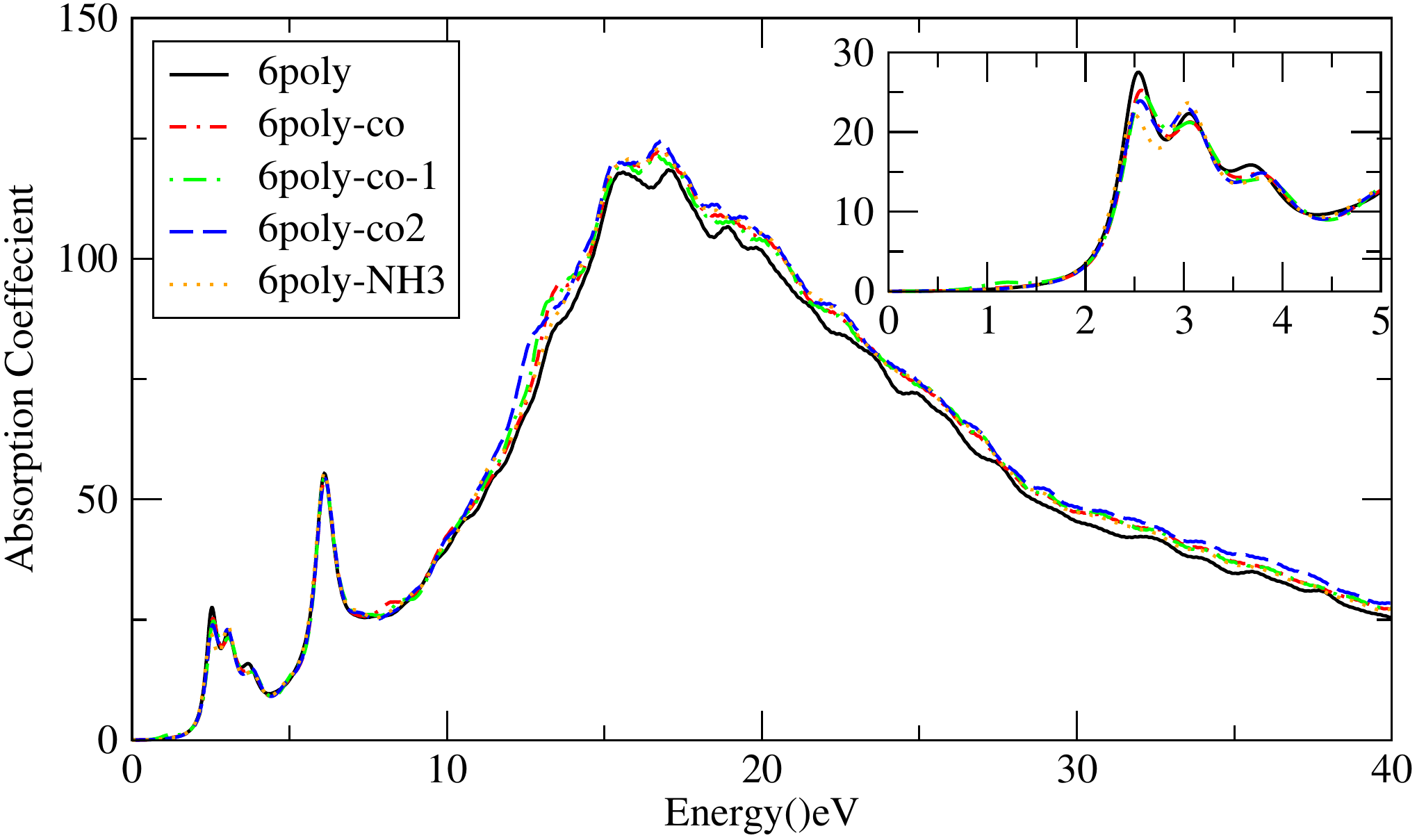}
\caption{\label{optic}
  Achieved absorption optical spectrum complexes 
  and transition states corresponding to nPANIES molecule.
  The insets present the initial part of the spectra in a slim window.
}
\end{figure}

The first peak of the absorption spectrum shows the first excited state and the energy that is 
equal to the energy needed for transfer of one electron from the (HOMO) to (LUMO).
The following peaks in the absorption spectrum represent the subsequent electrons 
excitation from other states. By enhancing the energy between different states, excitation 
happens in larger energy.
Fig.~\ref{optic} presents the optical spectrum complexes and the transition states corresponding
to nPANI ES molecule.
The first peak is just associated with (first excitation), 
but the following peaks are associated with several excitations which broadening for each peak affecting  by 
the amount of electron excitations in the corresponding energy.
The data that are calculated here showed in table \ref{r}.
We used Equation (6) to compute the adsorption energy of gas molecules on nPANI ES.

\begin{table}
\caption{\label{r}
The optical absorption spectrum factor for nPANI ES and its complexes using the Liovile-lanczos appraoch 
with TD-DFT.
$E_{F}$ :First excited energy.
$\lambda_{max}$ : maximum wave lenght, $\Delta$E : diffrence between 
first excited energy of nPANI ES with first excited 
energy of nPANI ES-X, where X=CO(1), CO(2), CO$_{2}$ and NH$_{3}$.
}
\begin{ruledtabular}
\begin{tabular}{ccccc}
structure   & $E_{F}$(eV)  &  $\lambda_{max}$(nm) & $\Delta$E(eV)  & Result  \\
\hline
Isolated 2PANI ES  & 3.60  & 344   &   -     &   -     \\
2PANI ES-CO(2)     & 3.57 & 348 & -0.02 & blue  \\
2PANI ES-CO(1)     & 3.56 & 348   & -0.04 & blue  \\
2PANI ES-CO$_{2}$  & 3.59 & 345 & -0.01 & blue   \\
2PANI ES-NH$_{3}$  & 3.50 & 353 & -0.09 & blue    \\
Isolated 4PANI ES   & 2.67 & 462 &    -    &   -     \\
4PANI ES-CO(2)     & 2.81 & 440 & 0.13   & red     \\
4PANI ES-CO(1)     & 2.81 & 440 & 0.13   & red   \\
4PANI ES-CO$_{2}$  & 2.81 & 440 & 0.13   & red   \\
4PANI ES-NH$_{3}$  & 2.77 & 446 & 0.09  & red    \\
Isolated 6PANI ES  & 2.54 & 487 &   -     &  -    \\
6PANI ES-CO(2)     & 2.57 & 482 & 0.02  & red   \\
6PANI ES-CO(1)     & 2.58  & 479 & 0.04  & red   \\
6PANI ES-CO$_{2}$  & 2.55 & 484 & 0.01  & red     \\
6PANI ES-NH$_{3}$  & 2.50 & 475 & -0.04 & blue   \\
\end{tabular}  
\end{ruledtabular}
\end{table}

The experimental $\lambda_{max}$  value of PANI ES is 356 nm (3.48 eV),\cite{vera2003synthesis} which is
0.12 eV lower than our calculated $\lambda_{max}$ value of 344 nm for free 2PANI ES (3.6 eV).
This difference is due to the infinite number of nPANI ES lengths
considered in emprical study while our computation simulated is on the basis of the limited length for 
a polymer having only two, four and six rings.

Obviously, from the data listed in table \ref{r}, when gas molecules interacts with nPANI ES,
the values for complexes are blue-shifted for 2PANI ES and its complexes, and red shifted for 
4PANI ES and 6PANI ES except 6PANI ES-NH${3}$ which are due to the $\pi$ $\longrightarrow$ $\pi^* $ 
transition. These value of blue and red shifts show 
the nPANI ES sensing ability toward CO(2), CO(1), CO$_{2}$ and NH$_{3}$. 
It can be said that more shifted value (but it is not true for all cases), 2PANI ES-NH$_{3}$ and
6PANI ES-NH$_{3}$ compared to other species which is totally in accordance
with their difference in the value of adsorption energy.

\section{CONCLUSION}

Density functional computations were carried out to study the structural, electronic, 
and optical properties of nPANI ES molecule and its complexes, with both QE and 
FHI-aims computational packages. The nearest distance between nPANI ES and gas molecules
is confirmed for NH$_{3}$ by the distance around of 1.8\AA{} which points to strong
interaction between them but CO(1) and CO(2) have the largest distance which
points to less interaction between them. This result is completely in accordance with
the obtained adsorption energies. The maximum and minimum decrease in angle $\angle$C$_{1}$N$_{2}$C$_{3}$
is about 2.07\AA{} and 1.1\AA{} in 4PANI ES-NH$_{3}$ and 6PANI ES-CO(1) complexes regarding their corresponding non-complexed nPANI ES. The larger
decrease in angle means the stronger electrostatic attraction between nPANI ES and NH$_{3}$. 
The kind of interaction between the tested species (CO, NH$_3$, CO$_2$)
with nPANI ES is studied with an understanding of the HOMO and
LUMO energies. simulating electronic features such as
HL gap from the energies of HOMO and LUMO supports the sensing ability of nPANI ES towards
the above-mentioned species.
The magnitude of Mulliken charge transfer is observed to be more when the adsorption of NH$_3$
molecule get adsorbed on nPANI ES comparing to the adsorption of other molecules.
The IR spectrum of the systems with adsorption of gas molecules on an isolated nPANI ES introduce
nPANI ES as a sensor. We can observe the amount of a higher and increased
IR intensity in 2PANI ES-NH$_{3}$, 4PANI ES-NH$_{3}$ and 6PANI ES-NH$_{3}$ complexes.
We computed optical properties of nPANI ES molecule and its complexes using 
optical spectrum, next compared them with experimental. The similarity of $\lambda_{max}$ and 
experimental $\lambda_{max}$ prove the effectiveness of Liouville-Lanczos 
method in computing the $\lambda_{max}$ and 
determining $\lambda_{max}$ . 
optical absorption spectrum analysis is also applied to all 
species and the resulted spectra indicates that the $\lambda_{max}$ were red or blue-
shifted depending on the type of complex (blue-shifted for 2PANI ES and its complexes, and 
red shifted for 4PANI ES and 6PANI-NH$_{3}$) because of $\pi$ $\longrightarrow$ $\pi^* $ 
transition which is can be considered as a testimony for the success of
interaction between nPANI ES and the species tested. 
All analyses reveal a physisorption procedure for all species over
interaction with nPANI ES.

\section{ACKNOWLEDGMENTS}

The authors thankfully appreciate the Sheikh Bahaei National High Performance Computing Center (SBNHPCC) for 
providing time and calculation facilities. SBNHPCC is supported with scientific and technological department of presidential 
office and Isfahan University of Technology (IUT). We want to appreciate Mr. Habib Ullah for the fine work he did on preparing structures.

\bibliography{LMO-STO}
\end{document}